\documentclass[12pt,twocolumn,twoside]{article}

\usepackage{titlesec}
\usepackage{fancyhdr}
\usepackage{authblk}
\usepackage{siunitx}
\usepackage{setspace,verbatim}
\usepackage{transparent}

\usepackage{xcolor,colortbl}
\usepackage{graphicx}
\usepackage{longtable}
\usepackage{booktabs}
\usepackage{multirow}
\usepackage[colorlinks, linkcolor=blue, anchorcolor=blue, citecolor=blue]{hyperref}
\usepackage{subcaption}
\usepackage{amsfonts}
\usepackage{amssymb}
\usepackage{amsmath}
\usepackage{lscape}
\usepackage{blindtext}
\usepackage{subcaption}
\usepackage{textcomp}
\usepackage{gensymb}
\usepackage{threeparttable}
\usepackage{booktabs, caption, makecell}
\usepackage{url}

\usepackage{breakurl}
\usepackage{hyperref}

\usepackage[top=1in,bottom=1in,left=1in,right=1in]{geometry}

\def\01{\ensuremath{0\mathord{-}1}}
\def\st{\mathop{\rm s.t.}}

\def\R{{\mathbb R}}

\usepackage[customcolors]{hf-tikz}

\tikzset{style green/.style={
    set fill color=green!50!lime!60,
    set border color=white,
  },
  style cyan/.style={
    set fill color=cyan!90!blue!60,
    set border color=white,
  },
  style orange/.style={
    set fill color=orange!80!red!60,
    set border color=white,
  },
  hor/.style={
    above left offset={-0.15,0.31},
    below right offset={0.15,-0.125},
    #1
  },
  ver/.style={
    above left offset={-0.1,0.3},
    below right offset={0.15,-0.15},
    #1
  }
}

\titleformat*{\section}{\normalsize\bfseries\sffamily}
\titleformat*{\subsection}{\small\bfseries\sffamily}
\titleformat*{\subsubsection}{\small\bfseries\sffamily}

\usepackage[font=small,labelfont=bf]{caption}

\newcommand\shorttitle{Optimization of circuiting arrangements for heat exchangers using DFO}
\newcommand\authors{Ploskas, Laughman, Raghunathan, and Sahinidis}

\fancypagestyle{firststyle}
{
   \fancyhf{}

   \fancyfoot[L]{\footnotesize\scshape \copyright 2017. This manuscript version is made available under the CC-BY-NC-ND 4.0 license \url{http://creativecommons.org/licenses/by-nc-nd/4.0/}.  The formal publication of this article is available at \url{https://doi.org/10.1016/j.cherd.2017.05.015}.}
}

\title{\textbf{Optimization of circuitry arrangements for heat exchangers using derivative-free optimization}}
\author[1]{Nikolaos~Ploskas}
\author[2]{Christopher~Laughman}
\author[2]{Arvind~U.~Raghunathan}
\author[1]{Nikolaos~V.~Sahinidis}
\affil[1]{Department of Chemical Engineering, Carnegie Mellon University, Pittsburgh, PA, USA}
\affil[2]{Mitsubishi Electric Research Laboratories, Cambridge, MA, USA}
\date{}

\begin{document}

\begin{singlespacing}
\twocolumn[{%
%%% title
\maketitle
\thispagestyle{firststyle}

% Define bibstyle
\bibliographystyle{plainnat}

%\doublespacing
%\baselinestretch

%\section**{Abstract}
%\vspace{-5pt}
%\pagenumbering{gobble}
\footnotesize
\vspace{-1cm}
\begin{abstract}
Optimization of the refrigerant circuitry can improve a heat exchanger's performance. Design engineers currently choose the refrigerant circuitry according to their experience and heat exchanger simulations. However, the design of an optimized refrigerant circuitry is difficult. The number of refrigerant circuitry candidates is enormous. Therefore, exhaustive search algorithms cannot be used and intelligent techniques must be developed to explore the solution space efficiently. In this paper, we formulate refrigerant circuitry design as a binary constrained optimization problem. We use CoilDesigner, a simulation and design tool of air to refrigerant heat exchangers, in order to simulate the performance of different refrigerant circuitry designs. We treat CoilDesigner as a black-box system since the exact relationship of the objective function with the decision variables is not explicit. Derivative-free optimization (DFO) algorithms are suitable for solving this black-box model since they do not require explicit functional representations of the objective function and the constraints. The aim of this paper is twofold. First, we compare four mixed-integer constrained DFO solvers and one box-bounded DFO solver and evaluate their ability to solve a difficult industrially relevant problem. Second, we demonstrate that the proposed formulation is suitable for optimizing the circuitry configuration of heat exchangers. We apply the DFO solvers to $17$ heat exchanger design problems. Results show that TOMLAB/glcDirect and TOMLAB/glcSolve can find optimal or near-optimal refrigerant circuitry designs after a relatively small number of circuit simulations.
\end{abstract}


\vspace*{0.8cm}

}]

\footnotesize
%\newpage

\section{Introduction}
\label{sec1}

Heat exchangers (HEXs) play a major role in the performance of many systems that serve prominent roles in our society, ranging from heating and air-conditioning systems used in residential and commercial applications, to plant operation for process industries.  While these components are manufactured in a startlingly wide array of shapes and configurations~\cite{hsb94}, one extremely common configuration used in heating and air-conditioning applications is that of the crossflow fin-and-tube type, in which a refrigerant flows through a set of pipes and moist air flows across a possibly enhanced surface on the other side of the pipe, allowing thermal energy to be transferred between the air and the refrigerant.

Performance improvement and optimization of these components can be pursued by evaluating a number of different metrics, based upon the requirements of their application and their specific use case; these include component material reduction, size reduction, manufacturing cost reduction, reduction of pumping power, maximization of heating or cooling capacity, or some combination of these objectives. While some of these metrics are reasonably straightforward in concept (e.g., cost and size reduction), the heat capacity is influenced by many parameters, including the geometry of the heat exchanger, the inlet conditions on the air-side (temperature, velocity, and humidity), and the inlet conditions on the refrigerant side (temperature, pressure, and mass flux).  The aggregate performance of the entire fin-tube heat exchanger can thus be often viewed as the aggregate performance of the collection of tubes.

Due to the prevalence and importance of these components, systematic optimization of heat exchanger design has been a long standing research topic~\cite{fm57,hkv82,jp92}.  Many proposed methods use analytical approaches to improve the performance of heat exchangers.  Heddenrich et al.~\cite{hkv82} proposed a model to optimize the design of an air-cooled heat exchanger for a user-defined tube arrangement, in which parameters such as tubes diameter, length, and fin spacing are optimized subject to a given heat transfer rate between air and water. They developed a software for the analysis of air-cooled heat exchangers and was coupled with a numerical optimization program.  Ragazzi~\cite{r95} developed a computer simulation tool of evaporators with zeotropic refrigerant mixtures to investigate the influence of the number of coil rows and tube diameter on the overall heat exchanger performance.  Reneaume et al.~\cite{rpn00} also proposed a tool for computer aided design of compact plate fin heat exchangers, which allows optimization of the fins, the core, and the distributor under user-defined design and operating constraints.  They formulated a nonlinear programming problem and solved it using a reduced Hessian successive quadratic programming algorithm.

The configuration of the connections between refrigerant tubes in a fin-and-tube heat exchanger, also referred to as the refrigerant circuitry, has a significant effect on the performance of the heat exchanger, and as such has been studied as a candidate optimization variable.  Because non-uniform air velocities across the heat exchanger face can result in different air-side heat transfer characteristics and uneven refrigerant distribution can result in different refrigerant-side heat transfer and pressure drop behavior, the specific path followed by the refrigerant through the heat exchanger as it evaporates can have a significant influence on many of the performance metrics of interest as demonstrated by~\cite{lwn01,mlvb04,wjlc99,yl00}.  These researchers have studied the effect of improving refrigerant circuitry, and have concluded that circuitry optimization is often more convenient and less expensive as compared with other performance optimization approaches, such as changing the fin and tube geometries.  The optimal refrigerant circuitry for one heat exchanger has also been found to be different from that of another heat exchanger~\cite{ccc02,dyk05}.

While current approaches for heat exchanger design often rely on design engineers to choose the circuitry configuration based upon their experience and the output of an enumerated set of simulations, the highly discontinuous and nonlinear relationship between the circuitry and the HEX performance motivates the study of systematic methods to identify optimized refrigerant circuitry design.  Such a problem is particularly challenging because of the size of the decision space; even a simple HEX with $N$ tubes, one inlet, one outlet, and no branches or merges will have $N!$ possible circuitry configurations, making exhaustive search algorithms insufficient for searching the entirety of the solution space.  Moreover, there is no guarantee that the engineering effort required to use expert knowledge to optimize the HEX circuitry manually will result in an optimal configuration, especially for larger coils; a systematic optimization method that is capable of determining an optimal configuration would have the dual benefits of providing a better HEX and freeing up engineering time.

A variety of sophisticated approaches have recently been proposed to construct optimized refrigerant circuitry designs. Liang et al.~\cite{lwn00} proposed a model that can be used to investigate the performance of a refrigerant circuitry through exergy destruction analysis. Domanski and Yashar~\cite{dykm04} developed an optimization system, called ISHED (Intelligent System for Heat Exchanger Design), for finding refrigerant circuitry designs that maximize the capacity of heat exchangers under given technical and environmental constraints. Experiments demonstrated the ability of this tool to generate circuitry architectures with capacities equal to or superior to those prepared manually~\cite{dyk05,yld15,ywkd12}, particularly for cases involving non-uniform air distribution~\cite{dy07}.  Wu et al.~\cite{wdwf08} also developed a genetic algorithm that constructs every possible refrigerant circuitry to find an optimal circuitry configuration.  Bendaoud et al. \cite{boz10} developed a FORTRAN program allowing them to study a large range of complex refrigerant circuit configurations. They performed simulations on an evaporator commonly employed in supermarkets, showing the effect of circuiting on operation and performance.  Lee et al.~\cite{lkj16} proposed a method for determining the optimal number of circuits for fin-tube heat exchangers.  Their results demonstrated that this method is useful in determining the optimal number of circuits and can be used to determine where to merge or diverge refrigerant circuits in order to improve the heat exchanger performance.

The aforementioned methods generally require either a significant amount of time to find the optimal refrigerant circuitry or produce a circuitry for which it is difficult to verify the practicality of its application.  Genetic algorithms also generate random circuitry designs that may not satisfy connectivity constraints; feasible random circuitry designs for a HEX with one inlet and one outlet are easy to generate, but most randomly generated solutions with multiple inlets and outlets will be infeasible.  Random operators, such as those used in conventional genetic algorithms, consequently may not lead to efficient search strategies or even feasible circuit layouts.  Domanski and Yashar~\cite{dykm04} were able to circumvent such problems by using domain knowledge-based operators, i.e., only perform changes that are deemed suitable according to domain-knowledge, and use a symbolic learning method for circuit optimization.  Such a unique set of domain knowledge-based operators and rules for the symbolic learning method that can find good solutions for different types of heat exchangers is not easy to define, however.  These methods also may not efficiently explore the solution search space, as some tube connections are fixed during the optimization process~\cite{wdwf08}.

One of the contributions of this paper is the presentation of heat exchanger circuitry optimization methods that generate feasible circuit designs without requiring extensive domain knowledge.  As a result, the proposed approach can be readily applied to different types of heat exchangers.  We incorporate only realistic manufacturing constraints to the optimization problem in a systematic way.  We formulate the refrigerant circuitry design problem as a binary constrained optimization problem, and use CoilDesigner \cite{jar06}, a steady-state simulation and design
tool for air to refrigerant heat exchangers, to simulate the performance of different refrigerant circuitry designs.  We treat CoilDesigner as a black-box system and apply derivative-free optimization (DFO) algorithms to optimize heat exchanger performance.  While the DFO literature has recently been attracting significant attention, it currently lacks systematic comparisons between mixed-integer constrained DFO algorithms on industrially-relevant problems~\cite{rs13}.  A primary contribution of this paper is to provide results from a systematic comparison of four different mixed-integer constrained DFO algorithms and a box-bounded DFO algorithm that are applied to optimize heat exchanger circuitry using two different thermal efficiency criteria. We also use constraint programming methods to verify the results of the DFO methods for small heat exchangers.

The remainder of this paper is organized as follows. In Section~\ref{sec2}, we present circuitry design principles of a heat exchanger. Section~\ref{sec3} describes the proposed formulation for optimizing the performance of heat exchangers. Section~\ref{sec4} details the DFO solvers that are used in this work. Section~\ref{sec5} presents the computational experiments on finding the best circuitry arrangements for $17$ heat exchangers. Conclusions from the research are presented in Section~\ref{sec6}.

\section{Heat exchanger circuitry}
\label{sec2}

In general, the performance of a given heat exchanger depends on a wide variety of system parameters and inputs, including materials (e.g., working fluids, HEX construction), coil geometry (e.g., tube geometry, find construction), operating conditions (e.g., inlet temperature or humidity, mass flow rate), and circuitry configuration \cite{opdo07,sar08}.  For a given application or set of use cases, many of these parameters are set early in the design phase by economic or manufacturing process requirements.  The circuitry configuration, in fact, is also strongly influenced by manufacturing and economic constraints; this imposes important limits on the size of the decision space.  For the purposes of this paper, we will assume that all geometric and inlet characteristics are fixed, and that the main problem of interest is that of identifying the location and number of inlet and outlet streams, as well as the circuitry configuration, for a given HEX construction.  This describes a very practically-oriented problem, in which a manufacturing engineer is handed a specific coil and asked to specify the circuitry that will optimize its performance according to some metric.

\begin{figure*}[t]
\centering
\includegraphics[width=\textwidth]{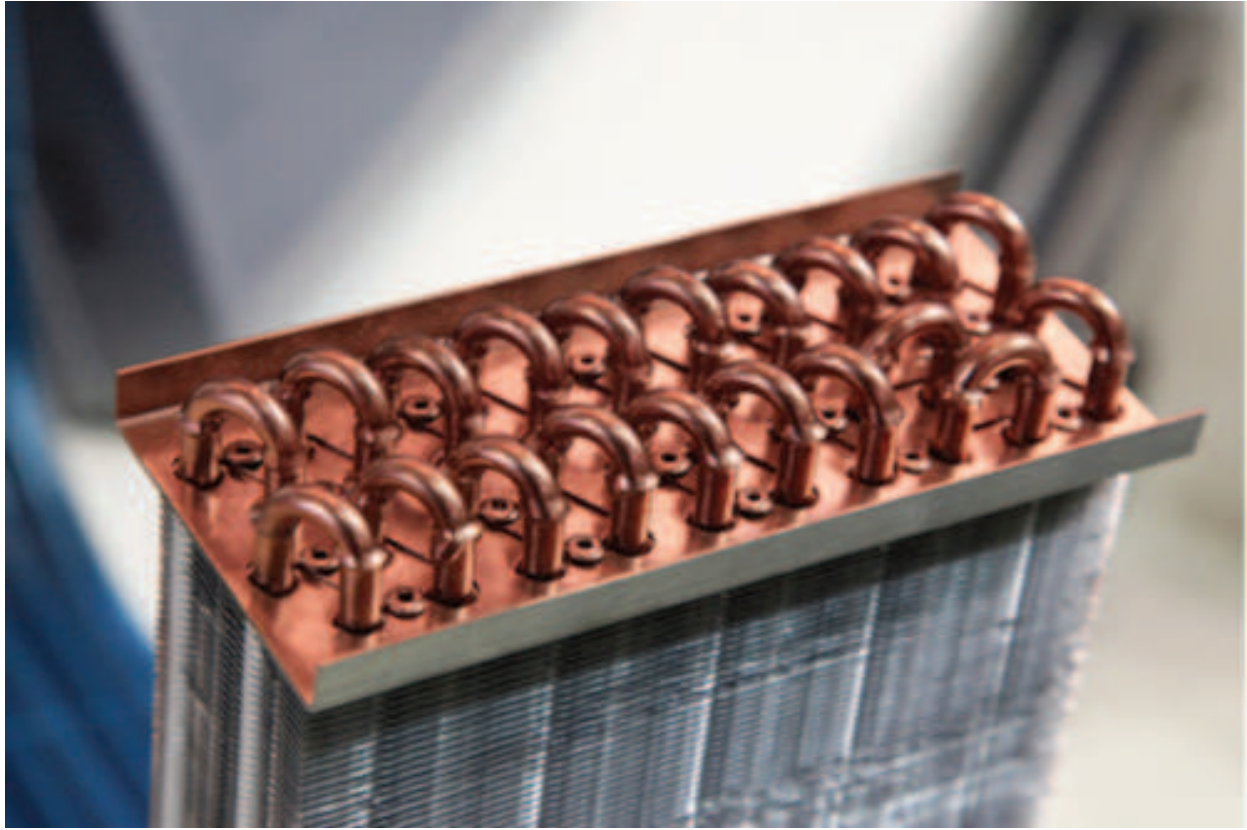}
\caption{Illustration of heat exchanger (Image licensed from S. S. Popov/Shutterstock.com)}
\label{fig1}
\end{figure*}

A picture illustrating the circuitry for a representative heat exchanger is illustrated in Figure~\ref{fig1}.  Such heat exchangers are typically constructed by first stacking layers of aluminum fins together that contain preformed holes, and then press-fitting copper tubes into each set of aligned holes.  The copper tubes are typically pre-bent into a U shape before insertion, so that two holes are filled at one time.  After all of the tubes are inserted into the set of aluminum fins, the heat exchanger is flipped over and the other ends of the copper tubes are connected in the desired circuitry pattern.  While the current picture only illustrates a very simple circuiting arrangement, many different connections can potentially be made between the tubes.

\begin{figure*}[t]
\centering
\includegraphics[width=0.4\textwidth]{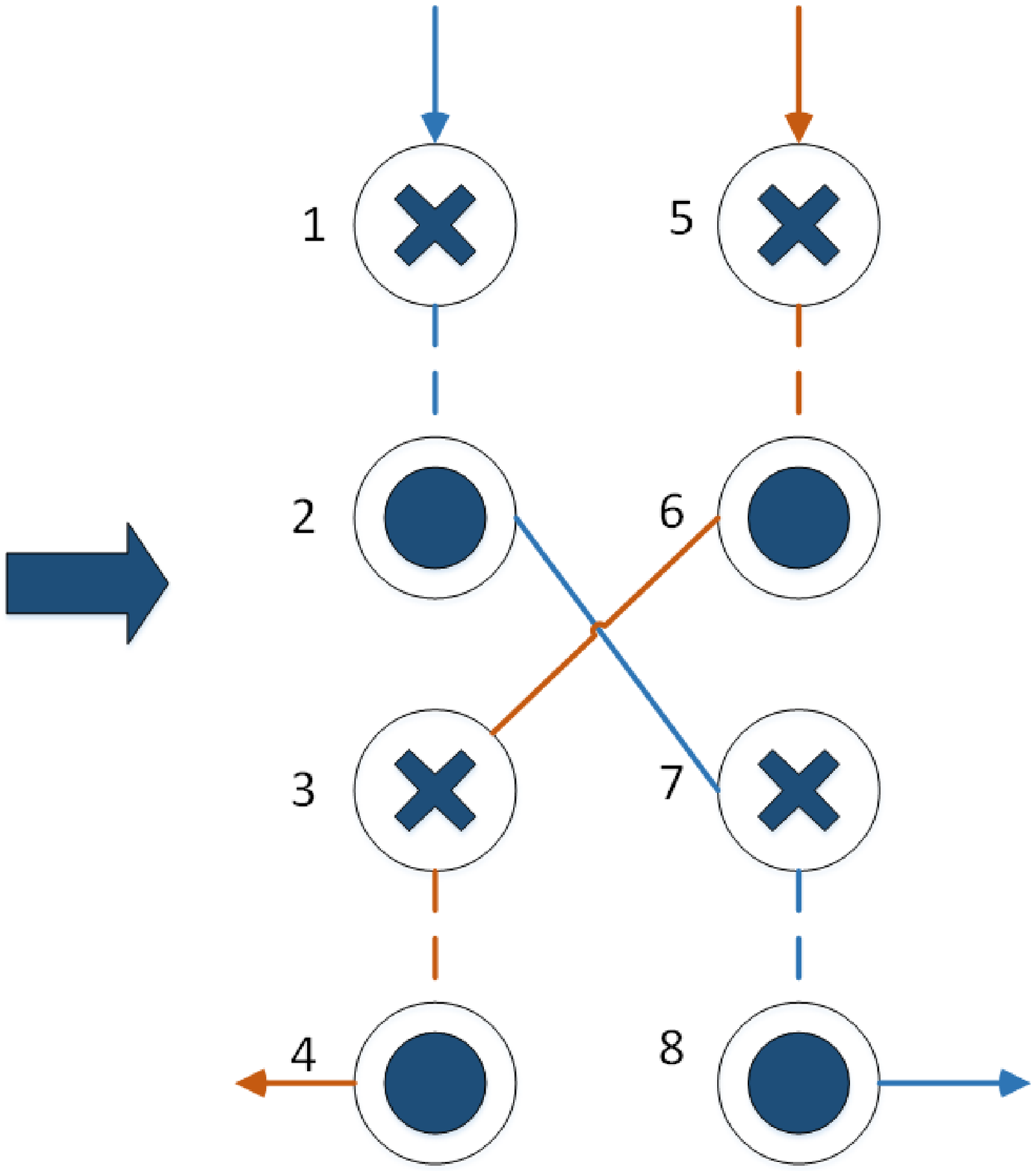}
\caption{Illustration of circuitry arrangement}
\label{cicruitry}
\begin{minipage}{0.9\textwidth}
\footnotesize
\emph{Notes: A crossed sign indicates that the refrigerant flows into the page, while a dotted sign indicates that the refrigerant flows out of the page. Different line colors are used to distinguish amongst different circuits.}
\end{minipage}
\end{figure*}

For the purposes of more clearly describing potential manufacturing constraints encountered in the construction of a fin-tube HEX, consider a diagram that illustrates the salient features relating to its circuitry.  Figure~\ref{cicruitry} illustrates a HEX constructed of eight tubes (each represented by a circle) with six connections of two types; one type of connection is at the far end of the tubes, while the other type of connection the near (front) end of the tubes.  In this framework, a crossed sign indicates that the refrigerant flows into the page, and a dotted sign indicates that the refrigerant flows out of the page.  Similarly, a dotted line between two tubes indicates a tube connection on the far end of the tubes, while a solid line indicates a tube connection on the front end of the tubes.  Different line colors are used to distinguish amongst different circuits.  Tubes are numbered in order of top to bottom in each row (normal to air flow), and left to right (in the direction of air flow). For the example figure, tubes $1$ and $5$ involve inlet streams, while tubes $4$ and $8$ involve outlet streams.

In light of this diagram, consider one set of manufacturing constraints imposed on the connections between tubes.  This set of constraints is such that adjacent pairs of tubes in each column, starting with the bottom tube, are always connected.  For example, in Figure~\ref{cicruitry}, this constraint implies that tubes $1$ and $2$, tubes $3$ and $4$, tubes $5$ and $6$, and tubes $7$ and $8$ are always connected.  The manufacturing process imposes this constraint because one set of bends at the far end of the coil are applied to the tubes before they are inserted into the fins, whereas the second set of connections or bends are introduced at the near end of the coil once a circuitry configuration is chosen.  Other related manufacturing-type restrictions used to constrain the space of possible circuiting configurations includes the following:
\begin{enumerate}
  \item Plugged tubes, i.e., tubes without connections, are not allowed
  \item The connections on the farther end cannot be across rows unless they are at the edge of the coil
  \item Inlets and outlets must always be located at the near end
  \item Merges and splits are not allowed.
\end{enumerate}

\begin{figure*}[t]
    \centering
    \begin{subfigure}[t]{0.5\textwidth}
        \centering
        \includegraphics[height=2.2in]{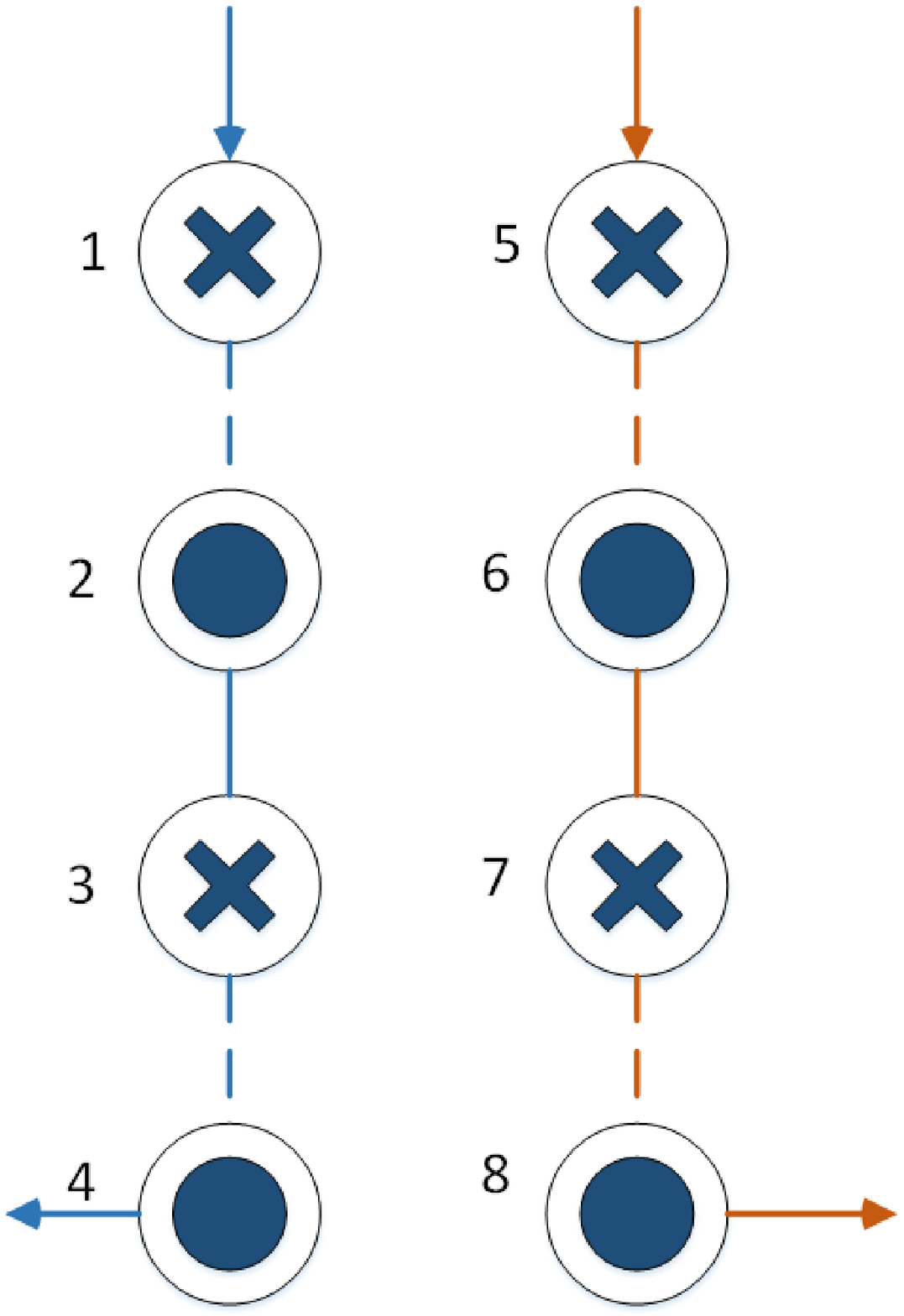}
        \caption{Valid circuitry arrangement}
        \label{validinvalidcircuitriesa}
    \end{subfigure}%
    ~
    \begin{subfigure}[t]{0.5\textwidth}
        \centering
        \includegraphics[height=2.2in]{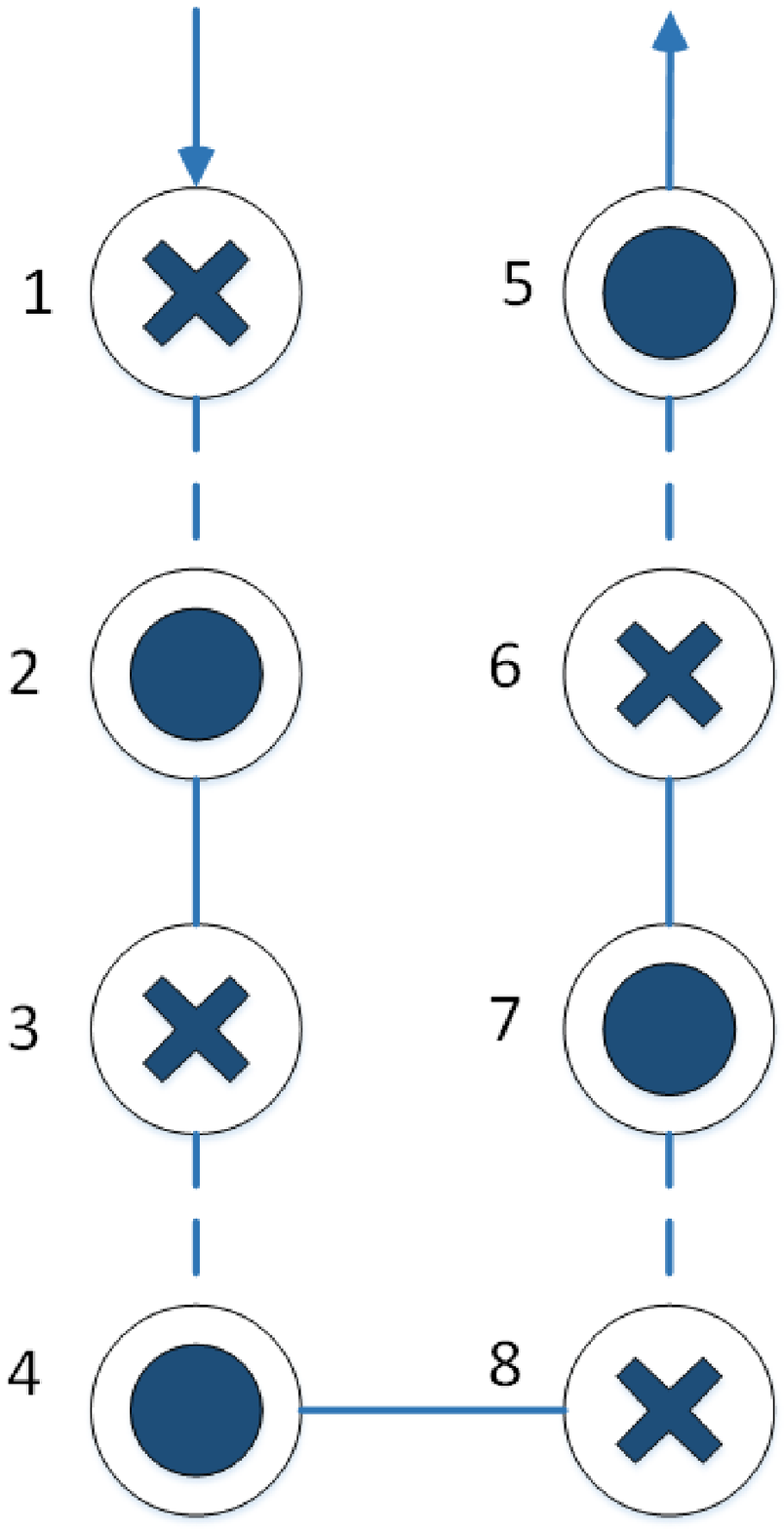}
        \caption{Valid circuitry arrangement}
        \label{validinvalidcircuitriesb}
    \end{subfigure}
    \\
    \begin{subfigure}[t]{0.5\textwidth}
        \centering
        \includegraphics[height=2.2in]{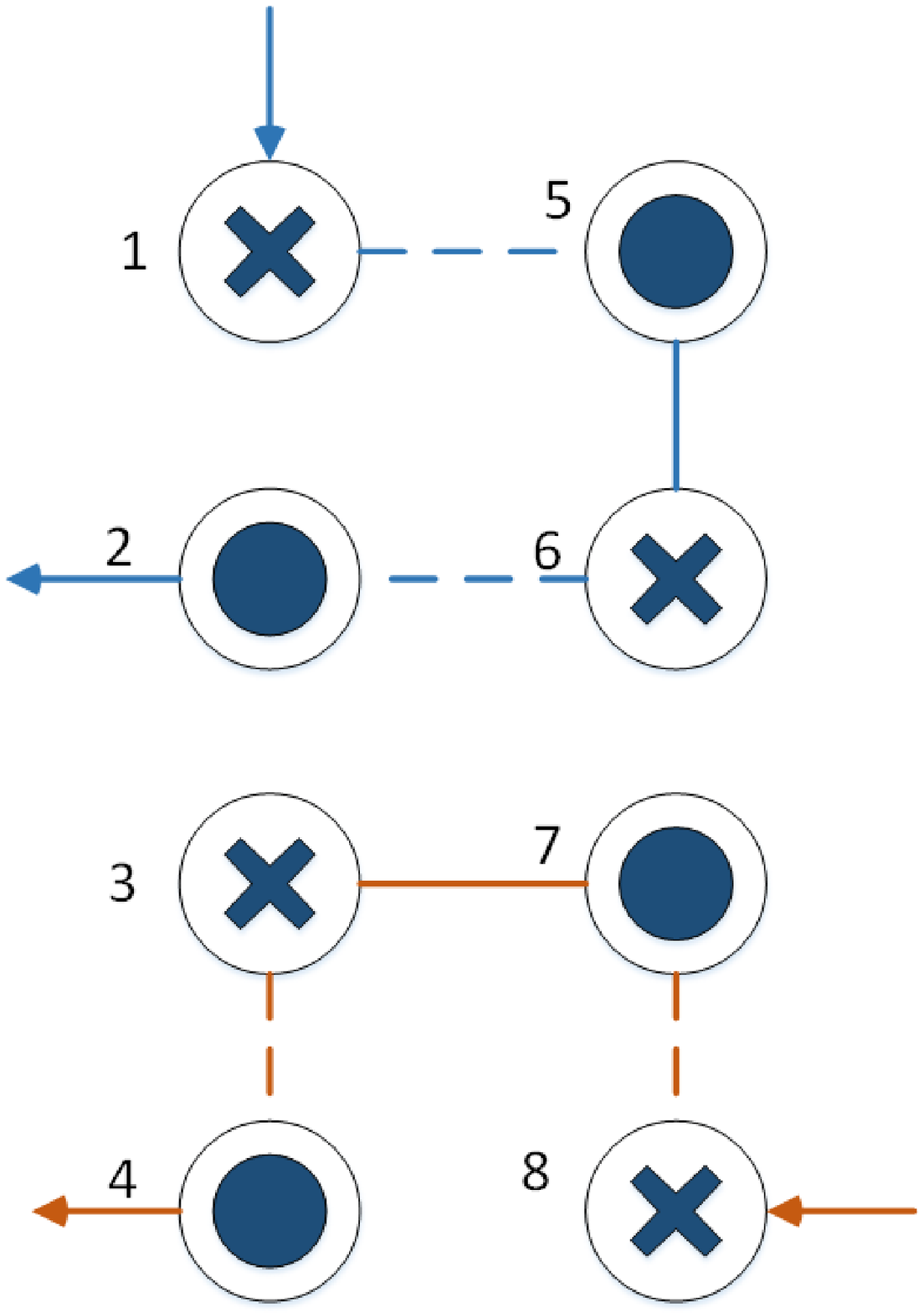}
        \caption{Invalid circuitry arrangement}
        \label{validinvalidcircuitriesc}
    \end{subfigure}%
    ~
    \begin{subfigure}[t]{0.5\textwidth}
        \centering
        \includegraphics[height=2.2in]{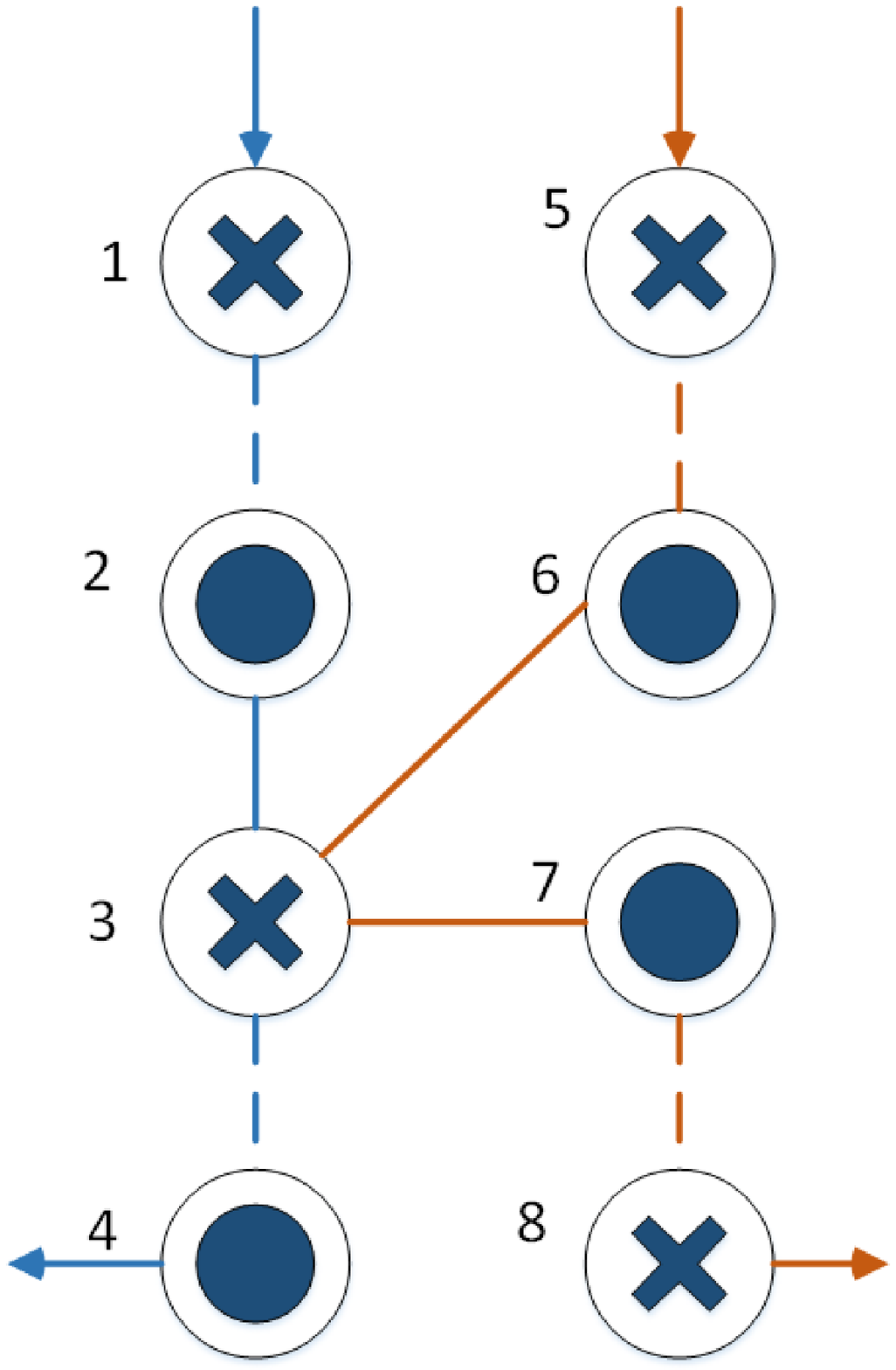}
        \caption{Invalid circuitry arrangement}
        \label{validinvalidcircuitriesd}
    \end{subfigure}
    \caption{Examples of valid and invalid circuiting arrangements}
    \label{validinvalidcircuitries}
\begin{minipage}{0.9\textwidth}
\footnotesize
\emph{Notes: A crossed sign indicates that the refrigerant flows into the page, while a dotted sign indicates that the refrigerant flows out of the page. Different line colors are used to distinguish amongst different circuits.}
\end{minipage}
\end{figure*}

Figure~\ref{validinvalidcircuitries} presents valid and invalid circuiting arrangements on a heat exchanger with eight tubes. The circuiting arrangement in Figure~\ref{validinvalidcircuitriesc} is not valid since it violates the second and third of the aforementioned restrictions, i.e., the connection between tubes $2$ and $6$ is not allowed and outlet tube $2$ is not located at the near end. In addition, the circuiting arrangement in Figure~\ref{validinvalidcircuitriesd} is invalid due to the merges and splits in tube $3$.

While this set of constraints represents one set of relevant manufacturing concerns, it does not represent the totality of such issues.  Other constraints might be included, such as penalties on the distance between tubes or the number of circuits.  Such constraints might also be incorporated into an optimization method, but are not included here for the sake of algorithmic and computational simplicity.

\section{Proposed model}
\label{sec3}

\subsection{Problem representation}
\label{sec3_1}

The problem representation in terms of an optimization formulation is one of the key aspects of optimization approaches that determines the degree of their success. Here, the refrigerant circuitry problem is represented as a large-scale binary combinatorial problem.  We use graph theory concepts to depict a circuitry configuration as a graph, where the tubes are the nodes and the connections between tubes are the edges. For example, the adjacency matrix for the circuitry configuration shown in Figure~\ref{cicruitry} is the following:
\begin{center}
$\begin{bmatrix}
0 & 1 & 0 & 0 & 0 & 0 & 0 & 0\\
1 & 0 & 0 & 0 & 0 & 0 & 1 & 0\\
0 & 0 & 0 & 1 & 0 & 1 & 0 & 0\\
0 & 0 & 1 & 0 & 0 & 0 & 0 & 0\\
0 & 0 & 0 & 0 & 0 & 1 & 0 & 0\\
0 & 0 & 1 & 0 & 1 & 0 & 0 & 0\\
0 & 1 & 0 & 0 & 0 & 0 & 0 & 1\\
0 & 0 & 0 & 0 & 0 & 0 & 1 & 0
\end{bmatrix}$
\end{center}

\begin{figure*}[t]
    \centering
    \begin{subfigure}[t]{0.5\textwidth}
        \centering
        \includegraphics[height=2.2in]{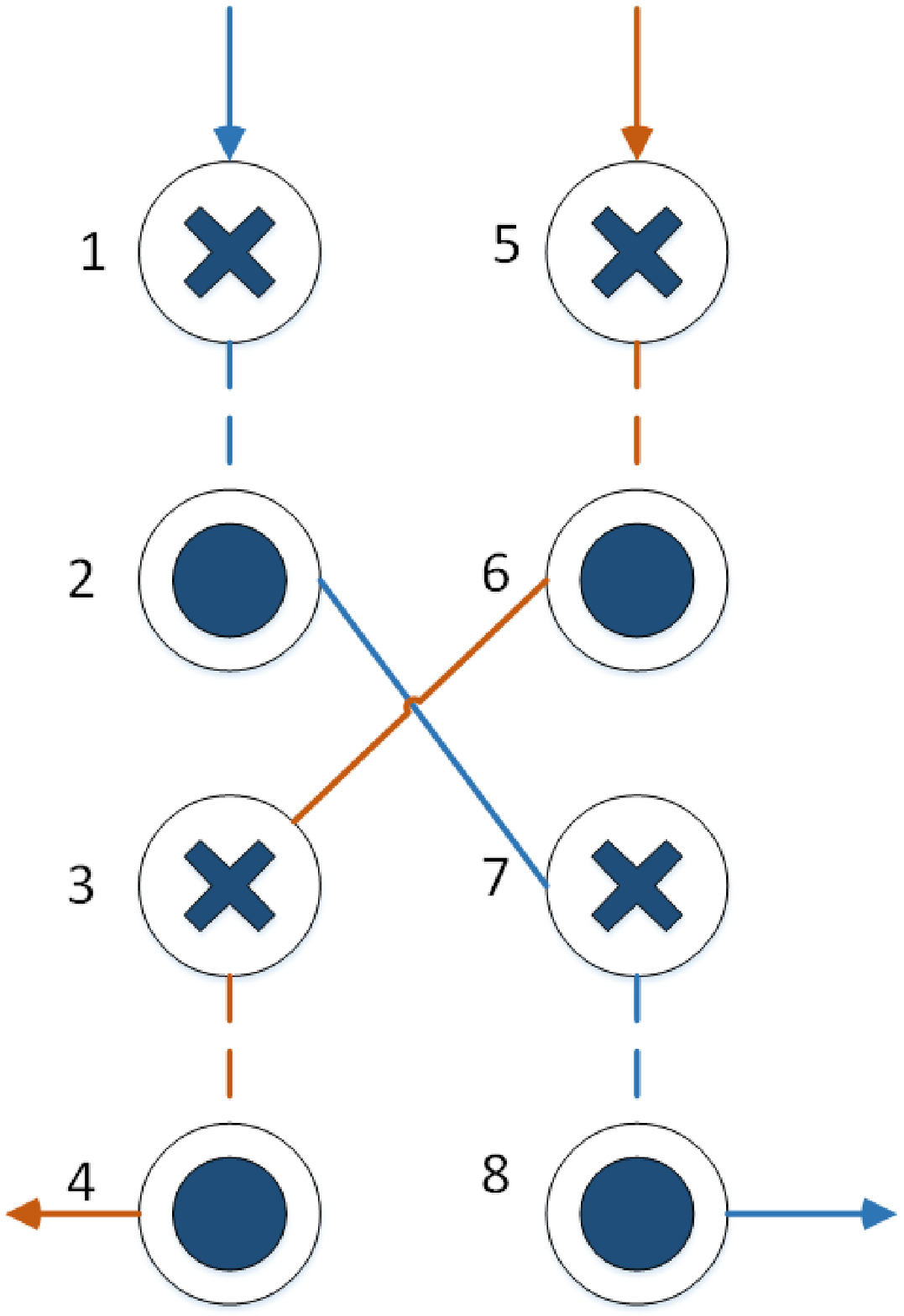}
        \caption{}
        \label{8tubescircuitriesa}
    \end{subfigure}%
    ~
    \begin{subfigure}[t]{0.5\textwidth}
        \centering
        \includegraphics[height=2.2in]{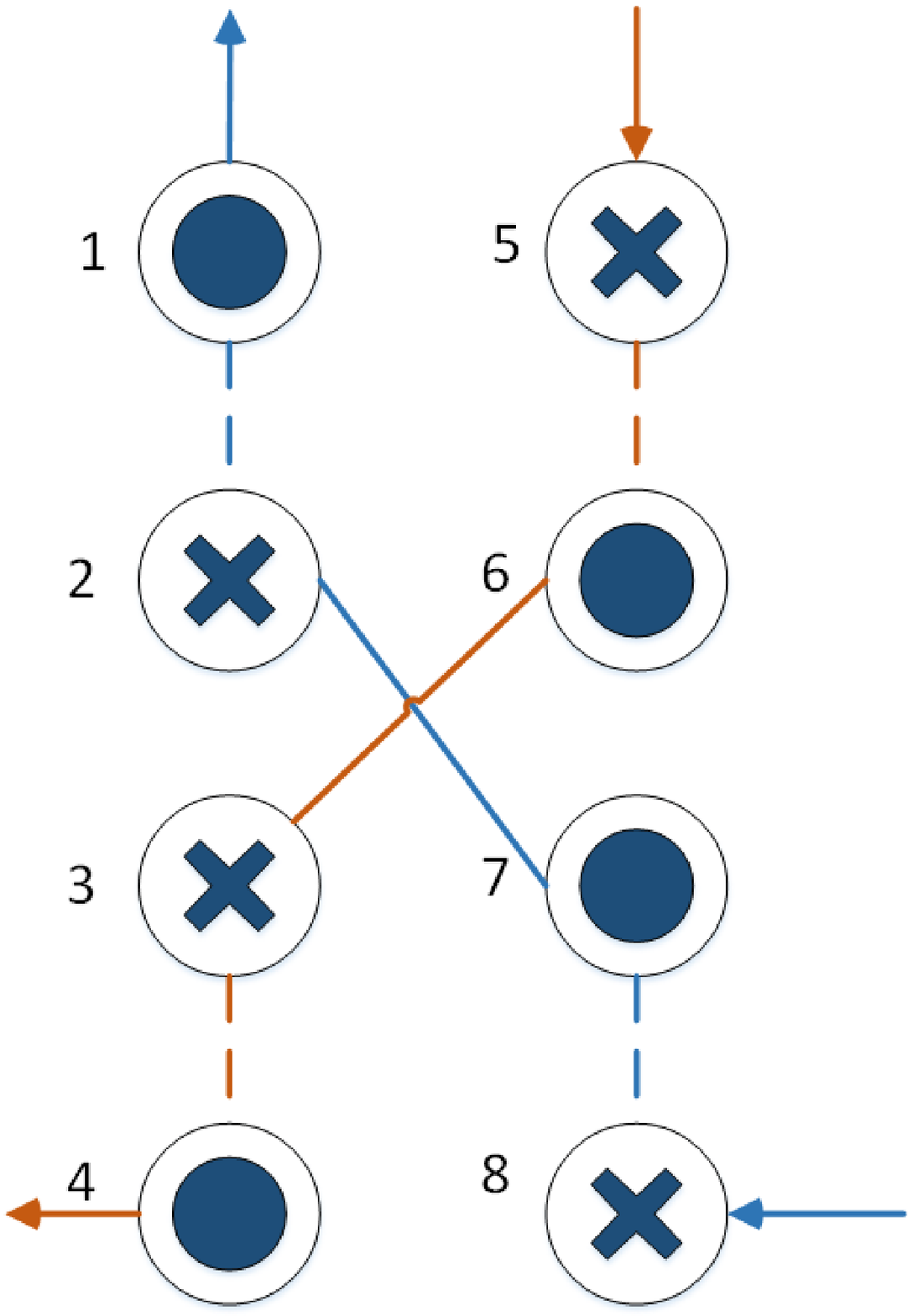}
        \caption{}
        \label{8tubescircuitriesb}
    \end{subfigure}
    \\
    \begin{subfigure}[t]{0.5\textwidth}
        \centering
        \includegraphics[height=2.2in]{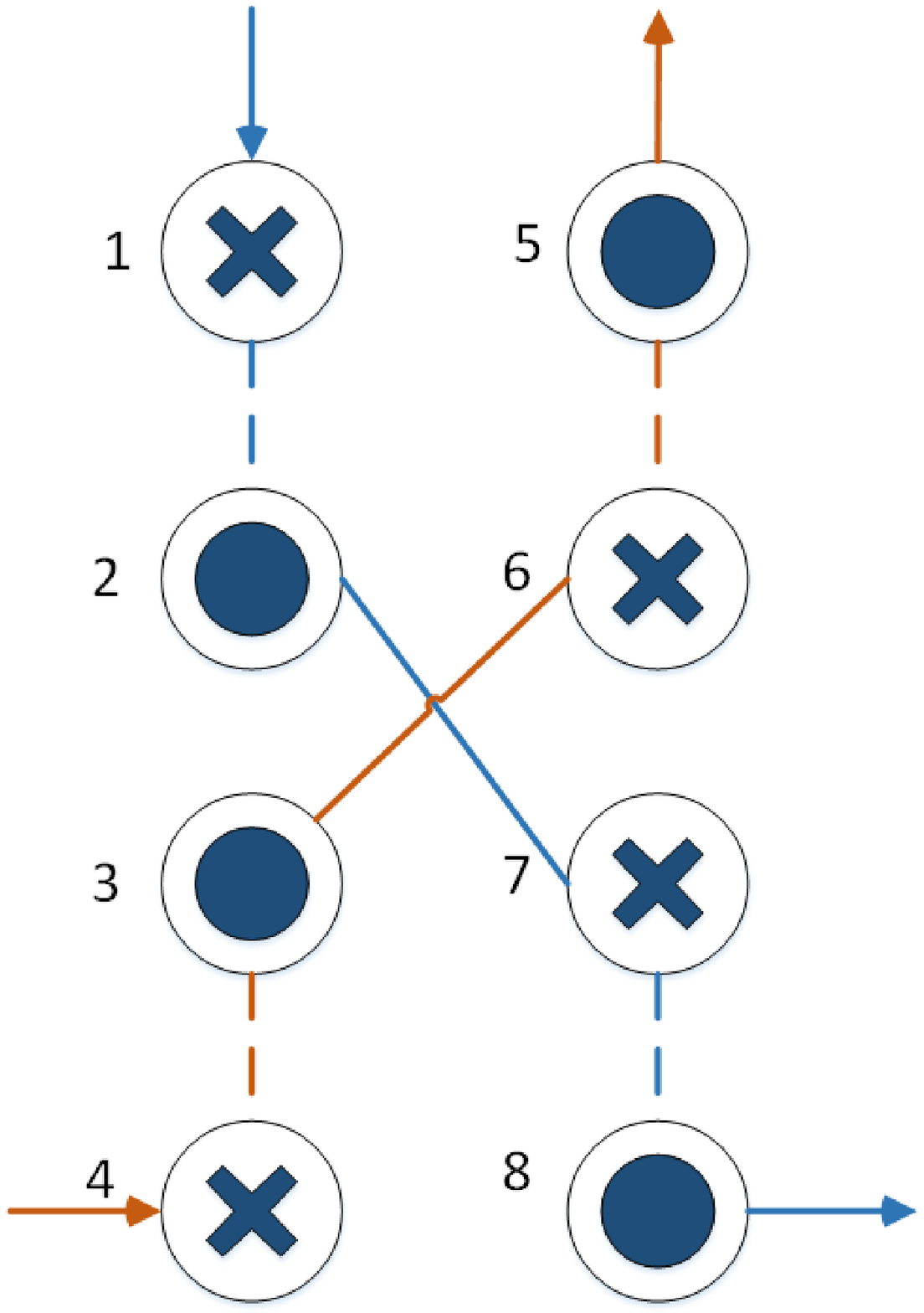}
        \caption{}
        \label{8tubescircuitriesc}
    \end{subfigure}%
    ~
    \begin{subfigure}[t]{0.5\textwidth}
        \centering
        \includegraphics[height=2.2in]{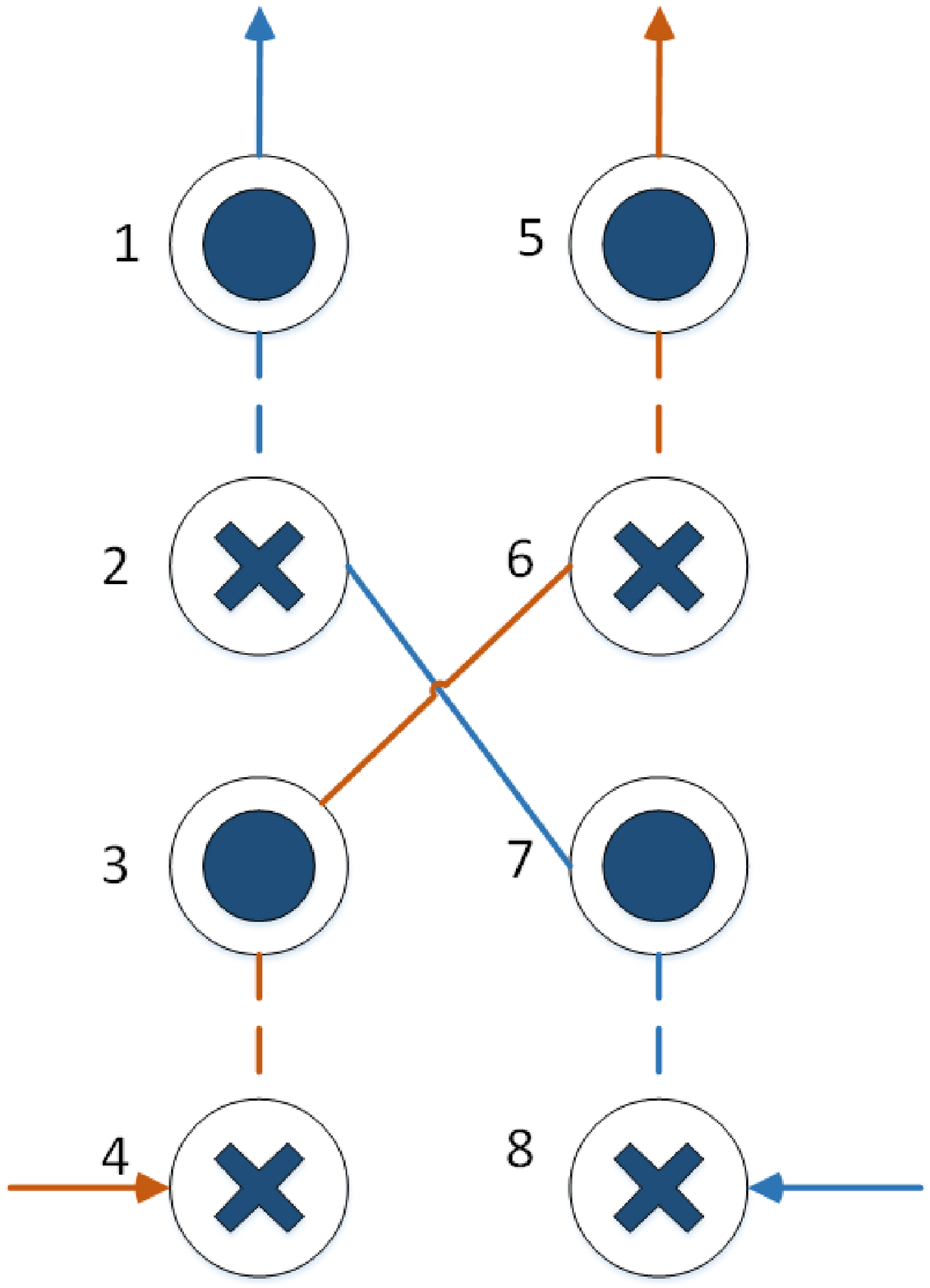}
        \caption{}
        \label{8tubescircuitriesd}
    \end{subfigure}
    \caption{Feasible circuitry designs for a heat exchanger with eight tubes and two circuits}
    \label{8tubescircuitries}
\begin{minipage}{0.9\textwidth}
\footnotesize
\emph{Notes: A crossed sign indicates that the refrigerant flows into the page, while a dotted sign indicates that the refrigerant flows out of the page. Different line colors are used to distinguish amongst different circuits.}
\end{minipage}
\end{figure*}

Binary variables will be used to model connections between tubes.  Since the graph is undirected, we need only the upper part of the adjacency matrix without the diagonal elements (no self-loops exist in a circuitry). Thus, we can limit the number of variables to $(t^2 - t)/2$, where $t$ is the number of tubes. The only drawback of treating the graph as undirected is that we do not know the start (inlet stream) and the end (outlet stream) of the circuits. Therefore, there are four feasible solutions for the above adjacency matrix (Figure~\ref{8tubescircuitries}). However, these feasible solutions produce very similar performance metrics. Extensive computational experiments showed that if a circuitry design has poor performance, it will not have a much better performance if we change the inlet and outlet streams. We preferred this approach, i.e., treating the graph as undirected, in order to create an optimization problem with significantly fewer variables, e.g., a heat exchanger with $36$ tubes can be modeled with only $630$ variables instead of $1,296$.

The vector of variables $x$ for the circuitry design problem contains $(t^2 - t)/2$ binary variables. Each variable is associated with the connection of two tubes. A variable $x_i$, $1 \leq i \leq (t^2 - t)/2$, is equal to~$1$ if the associated tubes are connected; otherwise $x_i = 0$. Let $Adj$ be the adjacency matrix. The elements of the solution vector $x$ are associated with an element of the upper part of matrix $Adj$ in order of left to right, and top to bottom:
\begin{center}
$\begin{bmatrix}
0 & x_1 & x_2 & \cdots & \cdots & x_{t-1}\\
0 & 0 & x_t & x_{t+1} & \cdots & x_{2t-3}\\
\vdots & \vdots & \vdots & \vdots & \cdots & \vdots\\
\vdots & \vdots & \vdots & \vdots & \cdots & \vdots\\
0 & 0 & 0 & \cdots & \cdots & x_{\left( t^2 - t \right) / 2}\\
0 & 0 & 0 & \cdots & \cdots & 0
\end{bmatrix}$
\end{center}

The adjacency matrix $Adj$ and the solution vector $x$ of the heat exchanger shown in Figure~\ref{cicruitry} are the following:
\begin{center}
$Adj = \begin{bmatrix}
- & 1 & 0 & 0 & 0 & 0 & 0 & 0\\
- & - & 0 & 0 & 0 & 0 & 1 & 0\\
- & - & - & 1 & 0 & 1 & 0 & 0\\
- & - & - & - & 0 & 0 & 0 & 0\\
- & - & - & - & - & 1 & 0 & 0\\
- & - & - & - & - & - & 0 & 0\\
- & - & - & - & - & - & - & 1\\
- & - & - & - & - & - & - & -
\end{bmatrix}$\\
$x^T = ( 1, 0, 0, 0, 0, 0, 0, 0, 0, 0, 0, 1, 0, 1, 0, 1, 0, $\\
$0, 0, 0, 0, 0, 1, 0, 0, 0, 0, 1 )$
\end{center}

\subsection{Objective function}
\label{sec3_2}

Various performance metrics have been used in order to evaluate and compare the performance of different circuitry designs~\cite{lkj16,wdwf08}. The most common goals when designing a heat exchanger is typically to maximize the heat capacity or to obtain the shortest joint tubes. Two targets of the refrigerant circuit optimization are considered in this work: (i) maximize the heat capacity, and (ii) maximize the ratio of the heat capacity to the pressure difference across the heat exchanger. Thus, the heat exchanger circuitry optimization problem can be symbolically expressed as:
\begin{enumerate}
  \item To maximize the heat capacity:
\begin{center}
$\begin{matrix}
\max & Q(x) & \\
\st & \text{constraints on the farther end} \\
     & \text{constraints on the front end} \\
 & x_i \in \left\{ 0, 1 \right\}, i = 1, 2, ..., n\\
\end{matrix}$
\end{center}
    where $Q$ is the heat capacity related to the solution vector $x$, $t$ is the number of tubes, $n = (t^2-t)/2$ is the number of decision variables, and the constraints on the farther and front end are presented in Section~\ref{sec3_3}.
  \item To maximize the ratio of the heat capacity to the pressure difference across the heat exchanger:
\begin{center}
$\begin{matrix}
\max & \frac{Q(x)}{\Delta P(x)} & \\
\st & Q(x) \geq Q_{lim}\\
     & \text{constraints on the farther end} \\
     & \text{constraints on the front end} \\
 & x_i \in \left\{ 0, 1 \right\}, i = 1, 2, ..., n\\
\end{matrix}$
\end{center}
    where $\Delta P$ is the pressure difference across the heat exchanger, and $Q_{lim}$ is a given limit for the heat capacity.
\end{enumerate}

\subsection{Constraints}
\label{sec3_3}

\begin{figure*}[t]
    \centering
    \begin{subfigure}[t]{0.5\textwidth}
        \centering
        \includegraphics[height=2.2in]{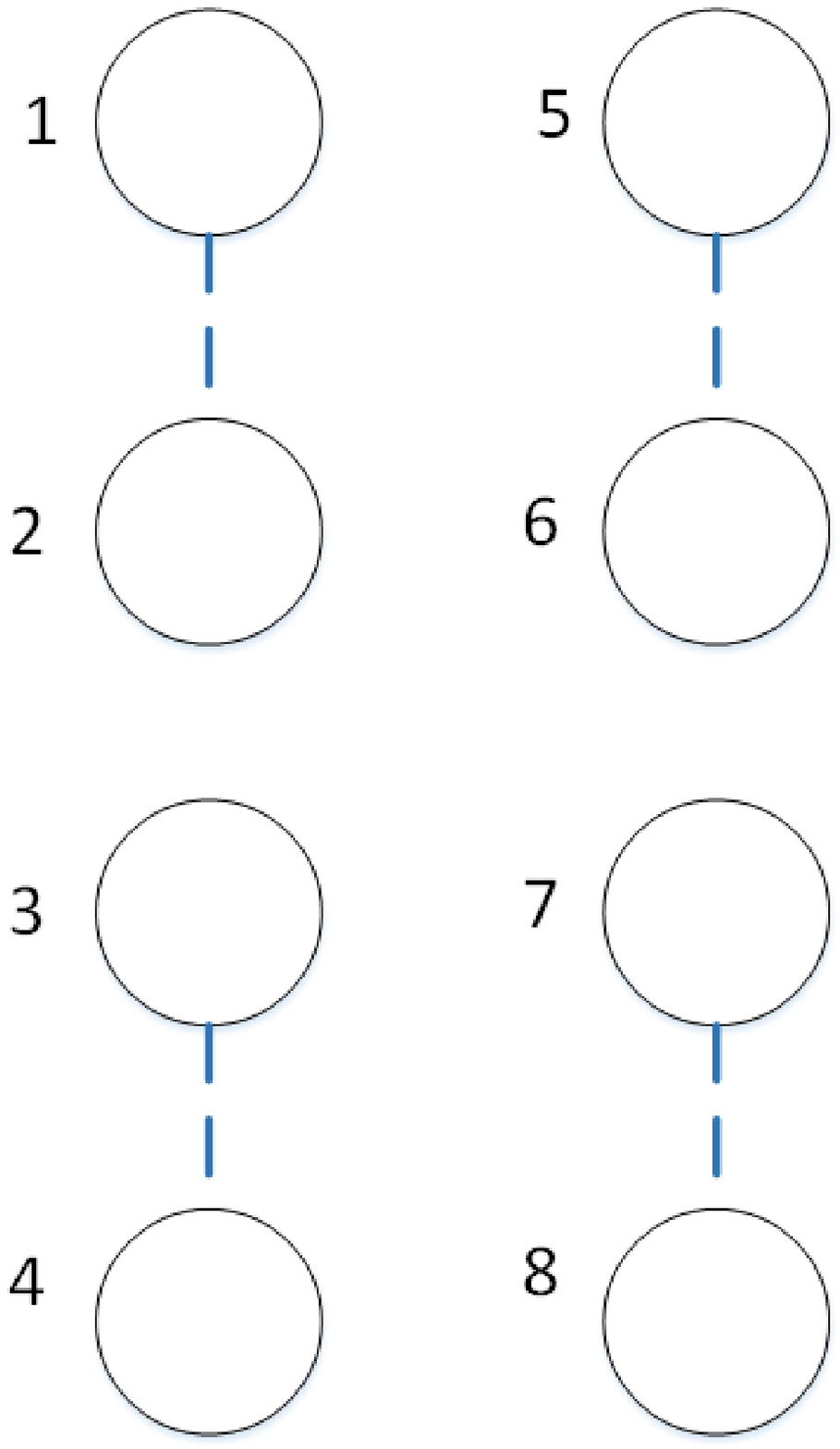}
        \caption{Heat exchanger with eight tubes}
        \label{tubesconnectionsa}
    \end{subfigure}%
    ~
    \begin{subfigure}[t]{0.5\textwidth}
        \centering
        \includegraphics[height=2.2in]{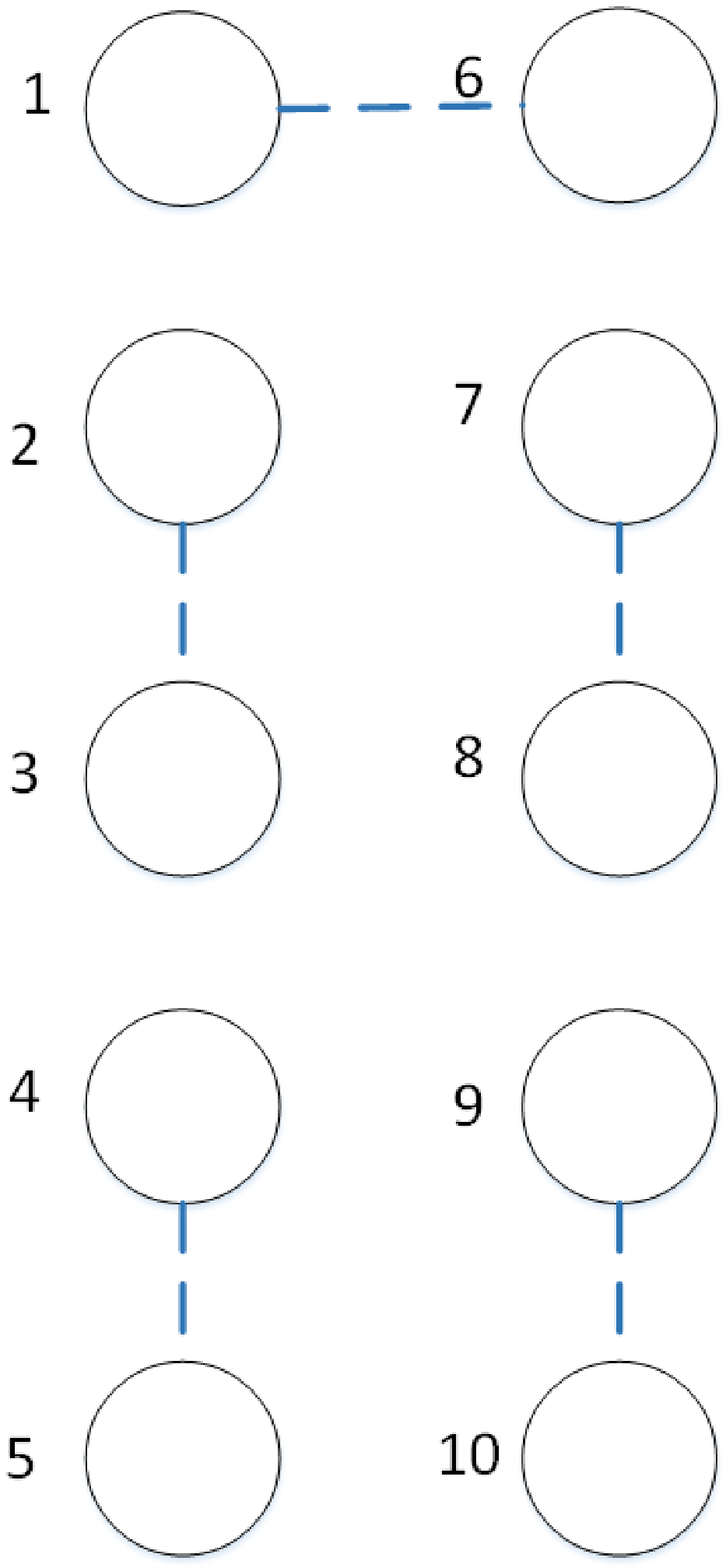}
        \caption{Heat exchanger with ten tubes}
        \label{tubesconnectionsb}
    \end{subfigure}
    \caption{Connections on the farther end}
    \label{tubesconnections}
\end{figure*}

As already discussed in Section~\ref{sec2}, there are two types of connections allowed, connections on the farther end of the tubes and connections on the front end of the tubes. In order to produce a feasible circuitry arrangement, some constraints are set. The constraints on the farther end are derived from the first two restrictions of the circuitry arrangement problem that were described in Section~\ref{sec2}: (i) plugged tubes are not allowed, and (ii) the connections on the farther end cannot be across rows unless they are at the edge of the coil. These two restrictions imply the constraints that should be set on the farther end. A heat exchanger with tubes in multiples of four has its tubes connected in pairs only in the same row; otherwise the first tubes in each row are connected together and the rest of the tubes are connected in pairs only in the same row. In each case, $t/2$ elements of vector $x$ are set equal to one.  Figure~\ref{tubesconnections} presents the connections on the farther end for a heat exchanger with eight tubes (Figure~\ref{tubesconnectionsa}) and for a heat exchanger with ten tubes (Figure~\ref{tubesconnectionsb}).

The restrictions on the connections on the front end are: (i) merges and splits are not allowed, and (ii) cycles are not allowed. The first restriction implies that every tube is connected with two tubes at most. Therefore, the sum of the elements of vector $x$ in each row $i$ and column $i$, $1 \leq i \leq n$, of the adjacency matrix should be less than or equal to two. The second restriction implies that we should avoid cycles when connecting tubes. We already have $t/2$ connections between tubes on the farther end. Hence, we should add a constraint for every combination of two, three, etc. pairs of these tubes in order not to form a cycle.

\subsection{Black-box model}
\label{sec3_4}

There are several simulation tools that have been developed for design and rating of heat exchangers like HTFS~\cite{aspen}, EVAP-COND~\cite{evap-cond}, and CoilDesigner~\cite{coildesigner}. We use the CoilDesigner to simulate the heat exchanger and compute the heat capacity and the ratio of the heat capacity to the pressure difference across the heat exchanger. We chose CoilDesigner for three reasons: (i) it is a highly customizable tool that allows the simulation of several types of heat exchangers, (ii) it has been validated on many data sets, and (iii) it provides an external communication interface for .NET framework. The existence of the external communication interface facilitates experimentation with different system parameters. The external interface also allows optimization studies to be carried out. In this study, we use the external communication interface to experiment with different designs and optimization algorithms in an entirely automated manner. Without such an interface, it would be impossible to perform the computational experiments in a reasonable amount of time through a graphical user interface of a simulation tool.

The exact relationship of the objective function with the decision variables is not explicit. CoilDesigner acts as a black-box model since we cannot deduce any explicit expression for the objective function. Hence, we can give as input to CoilDesigner the structural parameters and work conditions of a heat exchanger and receive as output many performance metrics about the function of the heat exchanger. A complete enumeration of all valid combinations is not possible for large heat exchangers. Thus, a more systematic and intelligent method should be utilized. Section~\ref{sec4} presents the DFO solvers that we used to solve this problem.

\section{Derivative-free optimization algorithms}
\label{sec4}

Derivative-free optimization or optimization over black-box models~\cite{rs13} is the optimization of a deterministic function $f: \R^n \rightarrow \R$ over a domain of interest that may include lower and upper bounds on the problem variables and/or general constraints. In typical DFO applications, derivative information is unavailable, unreliable, or prohibitively expensive. DFO has been a long standing research topic with applications that range from science problems to medical problems to engineering problems (see discussion and references in~\cite{rs13}).

Historically, the development of DFO algorithms started with the works of Spendley et al.~\cite{shh62} and Nelder and Mead~\cite{nm65}. Recent works on the subject offered significant advances by providing convergence proofs~\cite{aa06,csv09a,lt02}, incorporating the use of surrogate models~\cite{bmt04,serafini:98}, and offering software implementations of several DFO algorithms~\cite{nomad,dfo,cma-es}.

According to Rios and Sahinidis~\cite{rs13}, DFO algorithms can be classified as:
\begin{itemize}
  \item direct or model-based: direct algorithms determine search directions by computing values of the function $f$ directly, while model-based algorithms construct and utilize a surrogate model of the function $f$ to guide the search process
  \item local or global: depending upon whether they can refine the search domain arbitrarily or not
  \item stochastic or deterministic: depending upon whether they require random search steps or not
\end{itemize}

In this paper, we formulate the refrigerant circuitry design problem as a binary constrained optimization problem. Hence, DFO solvers that can handle constraints and discrete variables are preferred.  While the DFO literature has been attracting increasing attention, it currently lacks systematic comparisons between mixed-integer constrained DFO algorithms.  Rios and Sahinidis~\cite{rs13} presented a systematic comparison of the performance of several box-bounded DFO solvers.  There are review papers about algorithmic developments in constrained DFO solvers~\cite{bmf16,ds11,klt03}, but none of them presents a comparison across various constrained DFO solvers.  Clearly, there is a need to systematically compare constrained DFO solvers and evaluate their ability to solve industrially-relevant problems.

In this paper, we use five DFO algorithms: CMAES, MIDACO, NOMAD, TOMLAB/glcDirect, and TOMLAB/glcSolve. We included CMAES in this study because its performance was the best amongst all stochastic DFO solvers in the extensive computational study of~\cite{rs13}. We chose the other four solvers since they can handle general constraints and discrete variables. A brief description of each solver is given below:
\begin{enumerate}
  \item CMAES~\cite{cma-es}: Covariance Matrix Adaption Evolution Strategy (CMAES) is a stochastic global DFO solver that can handle bound constraints. It is a MATLAB implementation of a genetic algorithm for nonlinear optimization in continuous domain. The algorithm progresses by learning covariance matrices, which helps approach the optimum and reduce population sizes significantly. By sampling a multivariate normal distribution with zero mean and covariance matrix, CMAES generates a cluster of new sampling points leading to a better solution.
  \item MIDACO~\cite{midaco}: MIDACO is a stochastic global DFO solver that can handle bound and general constraints. It implements an ant colony optimization algorithm~\cite{seb09} with the oracle penalty method~\cite{sg10} for constrained handling. The implemented ant colony optimization algorithm is based on multi-kernel Gaussian probability density functions that generate samples of iterates.
  \item NOMAD~\cite{nomad}: Nonsmooth Optimization by Mesh Adaptive Direct Search (NOMAD) is a direct local DFO solver that can handle bound and general constraints. It is a C++ implementation of the MADS method~\cite{ad06} with different families of directions including GPS, LT-MADS, and OrthoMADS in its poll step. Three strategies are integrated into NOMAD: (i) extreme barrier, (ii) filter technique, and (iii) progressive barrier (PB). It also applies a genetic search strategy derived from Variable Neighborhood Search (VNS)~\cite{mh97} to escape from local optima in searching global minima.
  \item TOMLAB/glcDirect~\cite[pp.112-117]{tomlab}: TOMLAB/glcDirect is deterministic global solver that can handle bound and general constraints. It implements an improved version of Jones at al.~\cite{jps93} DIRECT algorithm (DIvide a hyperRECTangle), a deterministic sampling method for solving multivariate global optimization problems under bound constraints.
  \item TOMLAB/glcSolve~\cite[pp.118-122]{tomlab}: TOMLAB/glcSolve is a deterministic global solver that can handle bound and general constraints. It implements an improved version of Jones et al.~\cite{jps93} DIRECT algorithm.
\end{enumerate}

TOMLAB/glcDirect and TOMLAB/glcSolve can handle general constraints and always produce feasible solutions. MIDACO and NOMAD use penalty approaches for constrained handling. Hence, we should check if the constraints are violated prior to calling CoilDesigner. CMAES does not explicitly handle constraints. However, we can return a null value in order to indicate that the generated circuitry is not feasible.

In the cases that we maximize the ratio of the heat capacity to the pressure difference across the heat exchanger, a black-box constraint also exists, $Q(x) \geq Q_{lim}$, where $Q_{lim}$ is a given limit for the heat capacity (in the computational experiments of this paper, we set this number equal to~$3,900$). After calling CoilDesigner, we can export the heat capacity and penalize the objective function if $Q(x) \le Q_{lim}$:
\begin{equation}
f(x) - \lambda \max\left(0, Q_{lim} - Q(x)\right)^2
\end{equation}
where $\lambda$ is a user-defined weight for the violations (in the computational experiments of this paper, we set this number equal to~$10^6$, i.e., a value that is an order of magnitudes larger than the expected values of $f(x)$).

\section{Computational study}
\label{sec5}

In order to validate the proposed model, we performed a computational study with the aim of optimizing the heat capacity and the ratio of the heat capacity to the pressure difference across the heat exchanger. For this study, we started by manually designing $17$ different circuitry architectures. The structural parameters and work conditions of the $17$ test cases are shown in Table~\ref{parameters}. The only difference between the test cases is the number of tubes per row, ranging from $2$ to~$18$ that result in heat exchangers having from $4$ to~$36$ tubes.

\begin{table*}[]
  \centering
  \caption{Structural parameters and work conditions}
    \begin{tabular}{lc|lc}
    \hline
    \multicolumn{2}{c|}{Structural parameters} & \multicolumn{2}{c}{Work conditions} \\
    \hline
    \# of depth rows & 2     & Refrigerant type & R134a \\
    Tube length (mm) & 1,143 & Refrigerant temperature (\degree C) & 7 \\
    Tube inside diameter (mm) & 9.40 & Refrigerant pressure (kPa) & 350 \\
    Tube outside diameter (mm) & 10.06 & Refrigerant mass flow rate (kg/s) & 0.02 \\
    Tube thickness (mm) & 0.33 & Refrigerant mass quality & 0.15 \\
    Tube horizontal spacing (mm) & 19.05 & Air inlet pressure (kPa) & 101.325 \\
    Tube vertical spacing (mm) & 25.40 & Air inlet temperature (\degree C) & 24 \\
    Tube internal surface & Smooth & Air flow rate ($m^3/s$) & 2 \\
    Fin spacing (mm) & 1.17 & & \\
    Fins per inch & 20 & & \\
    Fin thickness (mm) & 0.10 & & \\
    Fin type & Louver & & \\
    Louver pitch (mm) & 2 & & \\
    Louver height (mm) & 1 & & \\
    \hline
    \end{tabular}
  \label{parameters}
\end{table*}

Prior to applying the DFO solvers to optimize the different heat exchangers, we performed a simulation for all combinations of heat exchangers with $4$, $6$, $8$, $10$, and $12$ tubes. We formulated the circuitry optimization problem as a Constraint Satisfaction Problem (CSP) using Choco solver~\cite{chocoSolver} in order to automate the procedure of finding all possible feasible circuitry designs. Choco is an open-source software that is used to formulate combinatorial problems in the form of CSPs and solve them with constraint programming techniques. The implemented search strategies of Choco produce all feasible solutions for each heat exchanger. We can evaluate each solution and gather various statistics that will help us to evaluate the performance of the DFO solvers. Note that we need to perform all combinations of inlet and outlet tubes for each solution since we used an undirected graph to represent the problem. Therefore, Choco will enumerate all feasible solutions and for each solution we need to consider all different combinations of inlet and outlet tubes. For example, if Choco finds the solution represented in Figure~\ref{8tubescircuitriesa}, then we need to simulate all four combinations (Figures~\ref{8tubescircuitriesa} to~\ref{8tubescircuitriesd}) of inlet and outlet tubes. Table~\ref{enumerationstats} presents the number of solutions, the number of combinations, the number of combinations whose heat capacity is greater than $3,900W$, and the execution time for simulating all of the circuitry designs of heat exchangers with $4$, $6$, $8$, $10$, and $12$ tubes. The execution time reported for the heat exchanger with $12$ tubes includes the simulation of only one combination for each solution.

\begin{table*}[]
  \centering
  \caption{Statistics of complete enumeration for heat exchangers with $4$ to~$12$ tubes}
  \begin{threeparttable}
    \begin{tabular}{c|cccc}
    \hline
    \# of tubes & \# of solutions & \# of combinations & \shortstack{\# of combinations\\($Q \ge 3,900$)} & \shortstack{Execution time \\ (sec)} \\
    \hline
    4     & 5     & 12    & 2     & 4 \\
    6     & 37    & 104   & 48    & 72 \\
    8     & 361   & 1168  & 544   & 926 \\
    10    & 3,965  & 14,976 & 6,981  & 17,261 \\
    12    & 54,539 & 232,512 & 41,899 & 72,985 \\
    \hline
    \end{tabular}
    \begin{tablenotes}\footnotesize
    \item[] \emph{Notes: Solutions represent circuitries where the inlet and outlet tubes are not known. Different combinations of inlet and outlet tubes are simulated for each solution.}
    \end{tablenotes}
    \end{threeparttable}
  \label{enumerationstats}
\end{table*}

\begin{figure*}[t]
\centering
\includegraphics[width=\textwidth]{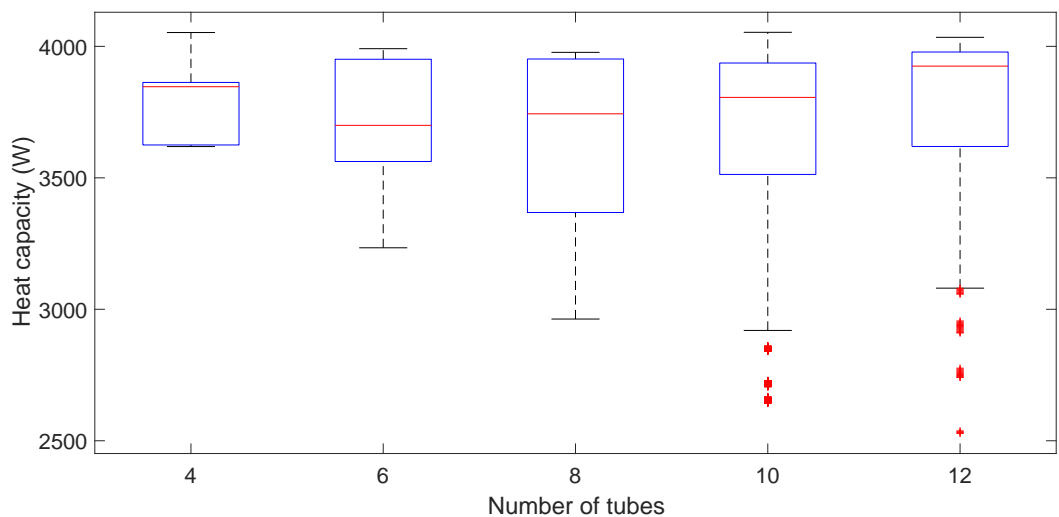}
\caption{Distribution of heat capacity for heat exchangers with $4$ to~$12$ tubes}
\label{heatcapacitydist}
\end{figure*}

\begin{figure*}[t]
\centering
\includegraphics[width=\textwidth]{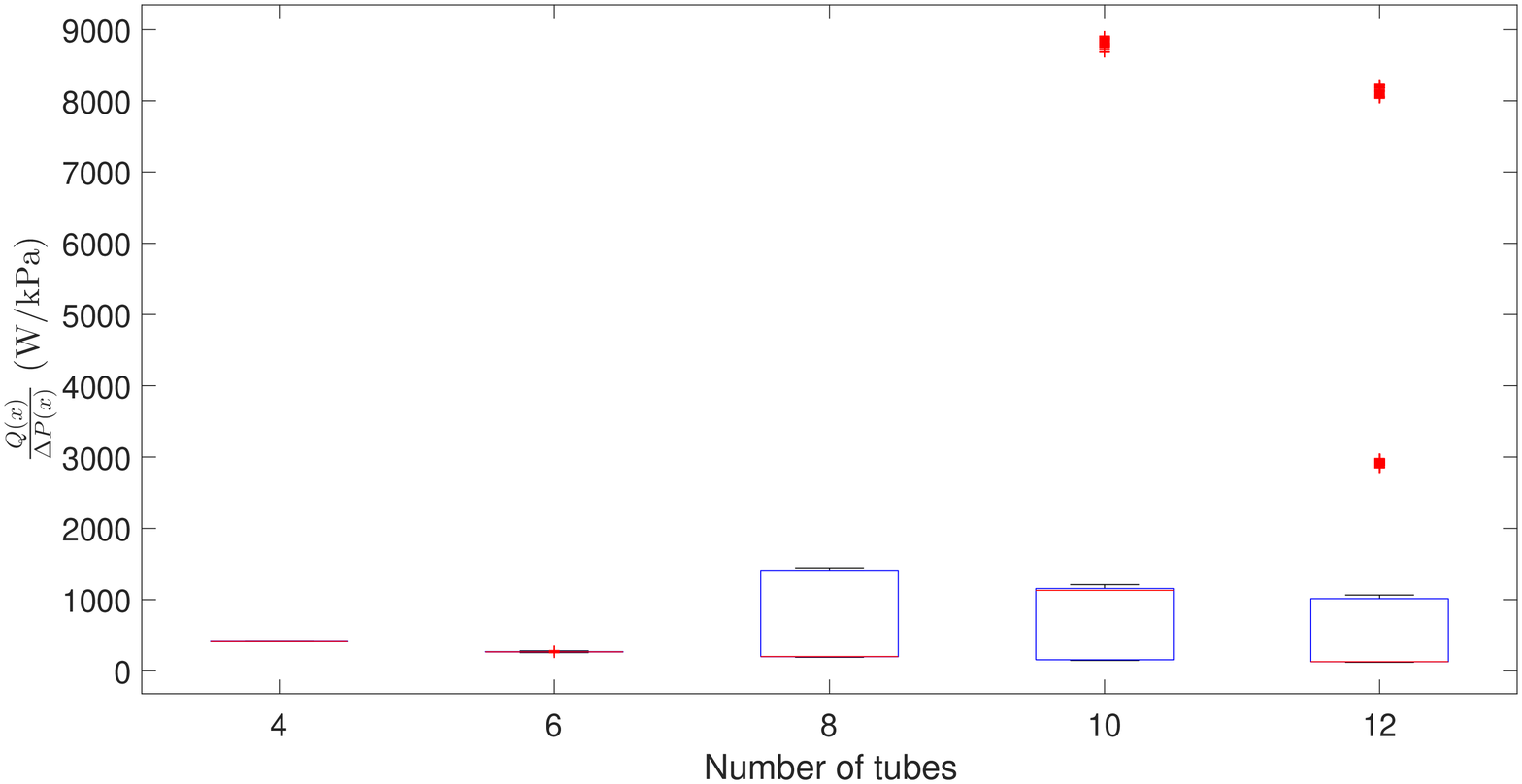}
\caption{Distribution of $Q(x)/\Delta P(x)$ for heat exchangers with $4$ to~$12$ tubes}
\label{pxdist}
\end{figure*}

The number of valid circuitry designs for a heat exchanger with $12$ tubes is $54,539$ and the total simulation time was ~$20$ hours. Hence, it is obvious from the results that the complete enumeration of all combinations is costly and time-consuming. However, the results of the complete enumeration will help us evaluate the performance of the DFO solvers in the next part of our computational experiments. Table~\ref{enumerationres} presents the results of the complete enumeration, while Figures~\ref{heatcapacitydist} and~\ref{pxdist} present the distribution of $Q$ and $Q(x)/\Delta P(x)$, respectively. For $Q(x)/\Delta P(x)$, we include only the combinations for which heat capacity is greater than $3,900W$. Results show that the optimal heat capacity is close to or above $4,000W$ for all heat exchangers. On the other hand, the optimal ratio of the heat capacity to the pressure difference across the heat exchanger ranges between $413W/kPa$ and $8906W/kPa$. The optimal solutions have objective function values that, on average, are $8$\% and $50$\% higher than the average heat capacity and pressure differences, respectively. Therefore, optimization of exchanger circuitry layout is very likely to improve significantly the efficiency of average heat exchanger designs. %xxx can we compare against a standard/naive approach? can we apply to literature examples (for larger problems) and compare against previous solutions? xxx

\begin{table*}[]
  \centering
  \caption{Results of complete enumeration for heat exchangers with $4$ to~$12$ tubes}
    \begin{tabular}{c|ccc|ccc}
    \hline
    \multirow{2}[0]{*}{\# of tubes} & \multicolumn{3}{c|}{$Q(x)$ ($W$)} & \multicolumn{3}{c}{$\frac{Q(x)}{\Delta P(x)}$ ($W/kPa$)} \\
          & Minimum & Maximum & Average & Minimum & Maximum & Average \\
    \hline
    4     & 3,619 & 4,053 & 3,807 & 407 & 413   & 410 \\
    6     & 3,234 & 3,991 & 3,700 & 254 & 280   & 268 \\
    8     & 2,963 & 3,977 & 3,675 & 190 & 1,446 & 560 \\
    10    & 2,643 & 4,053 & 3,649 & 147 & 8,906 & 775 \\
    12    & 2,528 & 4,034 & 3,716 & 120 & 8,229 & 575 \\
    \hline
    \end{tabular}
  \label{enumerationres}
\end{table*}

\begin{figure*}[t]
\centering
\includegraphics[width=\textwidth]{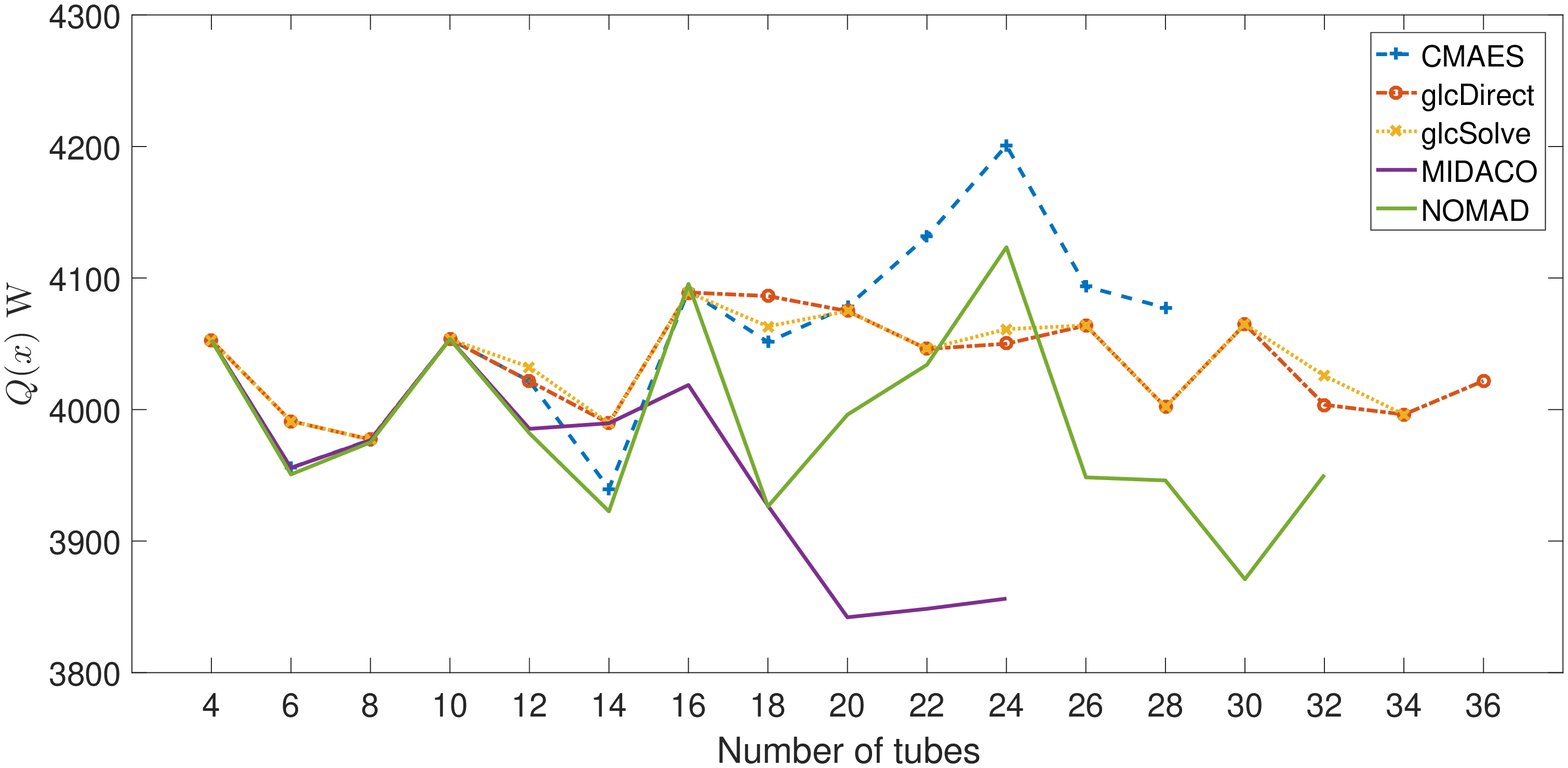}
\caption{Best solutions of heat capacity optimization}
\label{heatcapacitysol}
\end{figure*}

Next, we applied the five DFO solvers that were presented in Section~\ref{sec4} to the proposed constrained binary DFO problem. A limit of $2,500$ function evaluations and $86,400$ seconds was set for each run. Tables~\ref{heatcapacityresults1} to~\ref{pxresults2} present the detailed results of the optimization of the two objective functions, $Q(x)$ and $Q(x)/\Delta P(x)$. In each case, we report the best objective value, the execution time, and the number of function evaluations. A dash (``-'') is used to indicate when a solver did not find a feasible solution in the given limits. Figure~\ref{heatcapacitysol} presents a summary of the results for heat capacity optimization. TOMLAB/glcDirect and TOMLAB/glcSolve always find a solution that is optimal or near-optimal. TOMLAB/glcDirect and TOMLAB/glcSolve find the same solution on $12$ instances. TOMLAB/glcDirect finds a better solution for $18$ and $36$ tubes with heat capacities of $4,086$ $W$ and $4,022$ $W$, respectively, which represent a $0.57$\% and $2.16$\% improvement over TOMLAB/glcSolver. TOMLAB/glcSolve finds a better solution for $12$, $24$, and $32$ tubes with heat capacities of $4,032$ $W$, $4,061$ $W$, and $4,026$ $W$, respectively, which represent a $0.27$\%, $0.27$\%, and $0.55\%$ improvement over the results from TOMLAB/glcDirect.

CMAES performs well on most problems. It finds the best solution for $20$, $22$, $24$, $26$, and $28$ tubes with heat capacities of $4,078$ $W$, $4,132$ $W$, $4,201$ $W$, $4,094$ $W$, and $4,077$ $W$, respectively, which represent a $0.08$\%, $2.12$\%, $3.44$\%, $0.73$\%, and $1.89$\% improvement over the results from TOMLAB/glcSolve. However, it fails to solve the problems with more than $28$ tubes. MIDACO is able to find three best solutions for small heat exchangers ($4$, $10$, and $14$ tubes), but it fails to find a good solution for larger problems. In addition, MIDACO fails to even find a feasible solution for heat exchangers with more than $24$ tubes. Finally, the performance of NOMAD is not stable. It finds the best solution for $16$ tubes with a heat capacity of $4,095$ $W$, but it fails to solve the two largest problems.

Timewise, TOMLAB/glcSolve is faster than TOMLAB/glcDirect on smaller instances ($\leq 24$ tubes), but TOMLAB/glcDirect is much faster on larger instances ($\ge 24$ tubes) and on average. Moreover, TOMLAB/glcDirect and TOMLAB/glcSolve are faster than CMAES but slower than MIDACO and NOMAD. This was expected since MIDACO and NOMAD produce many infeasible solutions and CoilDesigner is not executed in such cases. Regarding the number of function evaluations, TOMLAB/glcSolve performs slightly better than TOMLAB/glcDirect, CMAES, and NOMAD, on average, while MIDACO always reaches the limit of function evaluations.

\begin{table*}[]
  \centering
  \caption{Computational results for heat capacity optimization--Part 1}
  \begin{threeparttable}
    \begin{tabular}{c|ccc|ccc}
    \toprule
    \multirow{2}[0]{*}{\# of tubes} & \multicolumn{3}{c|}{CMAES} & \multicolumn{3}{c}{glcDirect} \\
          & $Q(x)$ ($W$)     & Time  & \shortstack{Function\\evaluations} & $Q(x)$ ($W$)     & Time  & \shortstack{Function\\evaluations}  \\
    \hline
    4     & 4,053 &	66 &	164 & 4,053 & 5  & 5     \\
    6     & 3,956 &	142 & 206 & 3,991 & 38 & 37    \\
    8     & 3,977 &	332 & 457 & 3,977 & 531 & 361  \\
    10    & 4,053 &	928 & 932 & 4,053 & 1,171 & 509 \\
    12    & 4,022 &	2,624 &	1,474 & 4,022 & 1,862 & 533 \\
    14    & 3,940 &	8,468 &	2,059 & 3,990 & 1,575 & 438  \\
    16    & 4,090 &	14,720 &	2,500 & 4,089 & 5,518 & 1,306  \\
    18    & 4,051 &	16,666 &	2,500 & 4,086 & 5,469 & 952    \\
    20    & 4,078 &	7,927 &	2,500 & 4,075 & 5,535 & 886    \\
    22    & 4,132 &	18,380 &	2,500 & 4,046 & 7,404 & 945  \\
    24    & 4,201 &	29,301 &	2,500 & 4,050 & 18,688 & 1,309  \\
    26    & 4,094 &	41,529 &	2,500 & 4,064 & 19,631 & 1,261  \\
    28    & 4,077 &	63,046 &	2,500 & 4,002 & 23,833 & 1,489  \\
    30    &  - & - & - & 4,065 & 20,204 & 1,627  \\
    32    &  - & - & - & 4,004 & 33,781 & 1,309  \\
    34    &  - & - & - & 3,996 & 29,589 & 1,372  \\
    36    &  - & - & - & 4,022 & 39,201 & 1,512  \\
    \hline
    \shortstack{Geometric\\mean} & 4,055 & 4,309 & 1,287 & 4,034 & 3,590 & 578  \\
    \hline
    \end{tabular}
    \begin{tablenotes}\footnotesize
    \item[] \emph{Notes: A dash (``-'') is used to indicate when a solver did not find a feasible solution in the given limits.}
    \end{tablenotes}
    \end{threeparttable}
  \label{heatcapacityresults1}
\end{table*}

\begin{table*}[]
  \centering
  \caption{Computational results for heat capacity optimization--Part 2}
  \begin{threeparttable}
    \begin{tabular}{c|ccc|ccc}
    \toprule
    \multirow{2}[0]{*}{\# of tubes} & \multicolumn{3}{c|}{glcSolve} & \multicolumn{3}{c}{MIDACO}  \\
          & $Q(x)$ ($W$)     & Time  & \shortstack{Function\\evaluations} & $Q(x)$ ($W$)     & Time  & \shortstack{Function\\evaluations}  \\
    \hline
    4     & 4,053 & 4  & 5  & 4,053 & 1,403 & 2,500   \\
    6     & 3,991 & 35 & 37 & 3,956 & 545 & 2,500    \\
    8     & 3,977 & 384 & 361 & 3,977 & 385 & 2,500  \\
    10    & 4,053 & 820 & 481 & 4,053 & 577 & 2,500   \\
    12    & 4,032 & 1,122 & 472 & 3,985 & 329 & 2,500  \\
    14    & 3,990 & 1,553 & 442 & 3,990 & 407 & 2,500  \\
    16    & 4,089 & 5,294 & 921 & 4,019 & 1,093 & 2,500  \\
    18    & 4,063 & 5,254 & 840 & 3,927 & 1,303 & 2,500  \\
    20    & 4,075 & 5,456 & 890 & 3,842 & 2,272 & 2,500  \\
    22    & 4,046 & 7,324 & 946 & 3,849 & 3,026 & 2,500  \\
    24    & 4,061 & 17,150 & 1,431 & 3,856 & 3,938 & 2,500  \\
    26    & 4,064 & 20,316 & 1,261 & -     & -     & - \\
    28    & 4,002 & 26,528 & 1,496 & -     & -     & - \\
    30    & 4,065 & 22,683 & 1,287 & -     & -     & - \\
    32    & 4,026 & 46,540 & 1,267 & -     & -     & - \\
    34    & 3,996 & 75,810 & 1,376 & -     & -     & - \\
    36    & 3,935 &	86,400 & 985 & -     & -     & - \\
    \hline
    \shortstack{Geometric\\mean} & 4,030 & 3,740 & 537 & 3,954 & 988 & 2,500   \\
    \hline
    \end{tabular}
    \begin{tablenotes}\footnotesize
    \item[] \emph{Notes: A dash (``-'') is used to indicate when a solver did not find a feasible solution in the given limits.}
    \end{tablenotes}
    \end{threeparttable}
  \label{heatcapacityresults1}
\end{table*}

\begin{table*}[]
  \centering
  \caption{Computational results for heat capacity optimization--Part 3}
    \begin{threeparttable}
    \begin{tabular}{c|ccc}
    \toprule
    \multirow{2}[0]{*}{\# of tubes} & \multicolumn{3}{c}{NOMAD}  \\
          & $Q(x)$ ($W$)     & Time  & \shortstack{Function\\evaluations} \\
    \hline
    4     & 4,053	& 2.30 & 16 \\
    6     & 3,951 & 10 &	196 \\
    8     & 3,975 &	37 &	212 \\
    10    & 4,053 &	76 &	204 \\
    12    & 3,982 &	317 & 163 \\
    14    & 3,923 & 317 &	163 \\
    16    & 4,095 &	625 &	628 \\
    18    & 3,926	& 813 &	869 \\
    20    & 3,996	& 621 &	1,918 \\
    22    & 4,034	& 695 &	1,841 \\
    24    & 4,123	& 5,282 &	1,533 \\
    26    & 3,948 & 5,278 &	2,488 \\
    28    & 3,946 &	5,721 &	1,673 \\
    30    & 3,871 &	3,250 &	1,786 \\
    32    & 3,950 &	6,606 &	2,500 \\
    34    & - &	- &	- \\
    36    & - &	- &	- \\
    \hline
    \shortstack{Geometric\\mean} & 3,988 & 444 & 566 \\
    \hline
    \end{tabular}
    \begin{tablenotes}\footnotesize
    \item[] \emph{Notes: A dash (``-'') is used to indicate when a solver did not find a feasible solution in the given limits.}
    \end{tablenotes}
    \end{threeparttable}
  \label{heatcapacityresults2}
\end{table*}

\begin{figure*}[t]
\centering
\includegraphics[width=\textwidth]{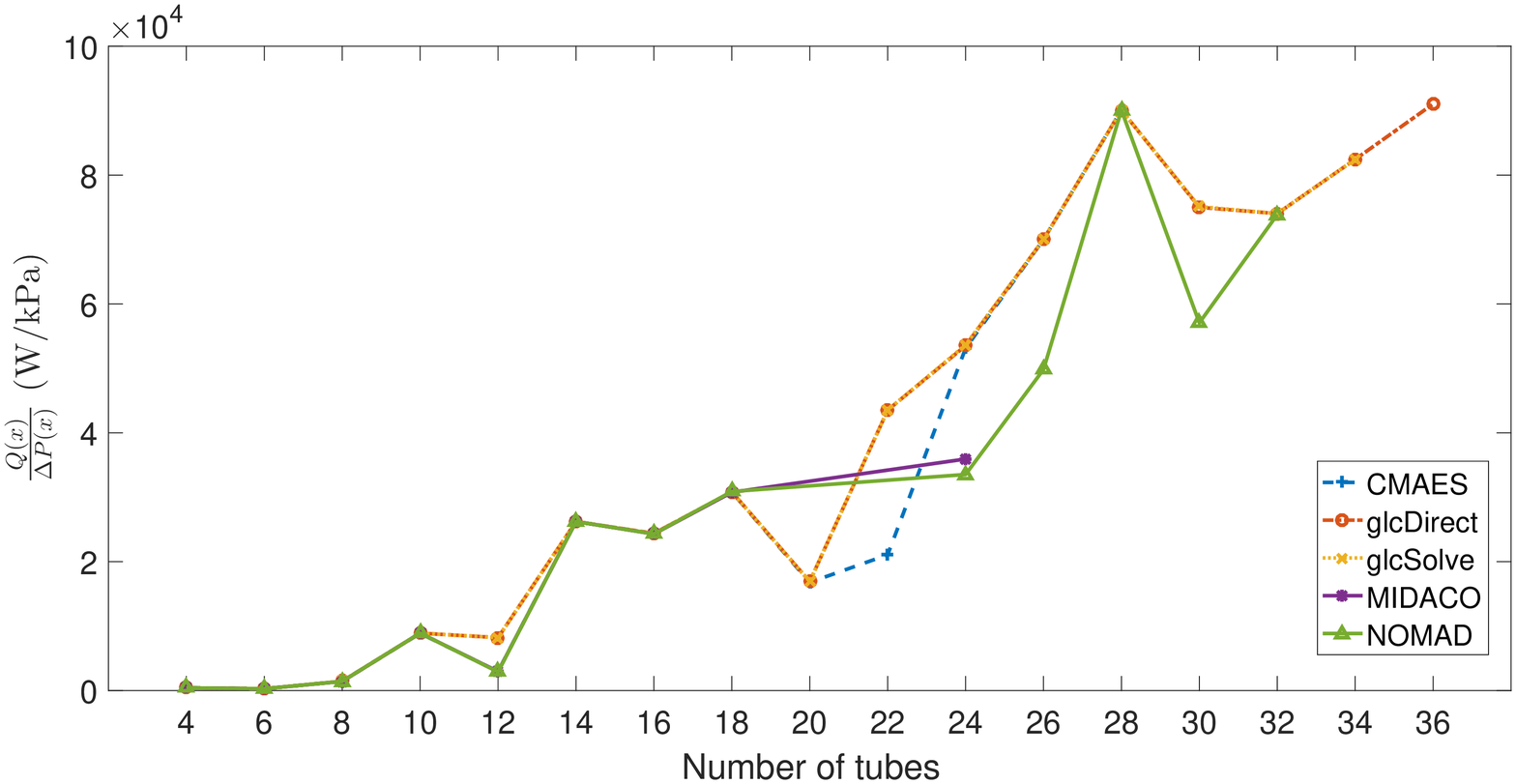}
\caption{Best solutions of $Q(x)/\Delta P(x)$ optimization}
\label{pxsol}
\end{figure*}

Figure~\ref{pxsol} presents a summary of the results for the optimization of the ratio of the heat capacity to the pressure difference across the heat exchanger. Similar to the results obtained for the optimization of the heat capacity, TOMLAB/glcDirect and TOMLAB/glcSolve always find a solution that is optimal or near-optimal. TOMLAB/glcDirect and TOMLAB/glcSolve find the same solution on $13$ instances. TOMLAB/glcDirect finds a better solution for $10$ tubes with an objective value of $8,900$ $W/kPa$, which represents a $0.07$\% improvement over TOMLAB/glcSolver. TOMLAB/glcSolve finds the best solution for $22$, $24$, and $30$ tubes with objective values of $43,517$ $W/kPa$, $53,646$ $W/kPa$, and $75,109/kPa$ $W$, respectively, which represent a $0.01$\%, $0.24$\%, and $0.15$ improvement over the results from TOMLAB/glcDirect.

CMAES performs well on most problems. It finds the best solution (along with other solvers) for $4$, $14$, and $28$ tubes. However, it fails to solve the problems with more than $28$ tubes. MIDACO is able to find some optimal solutions for small heat exchangers, but it fails to find a good solution for larger problems. In addition, MIDACO fails to find even a feasible solution for heat exchangers with $20$, $22$, and more than $24$ tubes. Finally, NOMAD performs well on most problems. It finds the best solution for $18$ tubes with an objective value of $30,889$ $W/kPa$, which represents a $0.19$\% improvement over TOMLAB/glcDirect and TOMLAB/glcSolve. It also finds the best solution (along with other solvers) on four other problems ($4$, $10$, $14$, and $18$ tubes).

Timewise, TOMLAB/glcSolve is faster than TOMLAB/glcDirect on smaller instances ($\leq 10$ tubes), but TOMLAB/glcDirect is much faster on larger instances ($\ge 10$ tubes), and on average. Moreover, TOMLAB/glcDirect and TOMLAB/glcSolve are faster than CMAES but slower than MIDACO and NOMAD. As already mentioned, MIDACO and NOMAD produce many infeasible solutions and CoilDesigner is not executed in such cases. Regarding the number of function evaluations, TOMLAB/glcSolve performs slightly better than TOMLAB/glcDirect on average. NOMAD performs less iterations than all other solver since it cannot solve the large problems. CMAES performs considerably more iterations than the aforementioned solvers, while MIDACO always reaches the limit of function evaluations.

\begin{table*}[]
  \centering
  \caption{Computational results for $Q(x)/\Delta P(x)$ optimization--Part 1}
    \begin{threeparttable}
    \begin{tabular}{c|ccc|ccc}
    \hline
    \multirow{2}[0]{*}{\# of tubes} & \multicolumn{3}{c|}{CMAES} & \multicolumn{3}{c}{glcDirect} \\
          & $\frac{Q(x)}{\Delta P(x)}$ ($W/kPa$     & Time  & \shortstack{Function\\evaluations} & $\frac{Q(x)}{\Delta P(x)}$ ($W/kPa$     & Time  & \shortstack{Function\\evaluations}  \\
    \hline
    4     & 413 &	35 &	74 & 413 & 4  & 5     \\
    6     & 277 &	168 &	374 & 280 & 38 & 37   \\
    8     & 1,432 &	243 &	704 & 1,443 & 422 & 361  \\
    10    & 8,881 &	2,424 &	1,262 & 8,900 & 1,143 & 518 \\
    12    & 2,941 &	258 &	1,314 & 8,216 & 1,435 & 653 \\
    14    & 26,219 &	453 &	2,500 & 26,219 & 2,920 & 774  \\
    16    & 24,348 &	10,330 &	2,500 & 24,393 & 10,822 & 1,289 \\
    18    & 30,803 &	850	& 2,500 & 30,830 & 7,525 & 1,100 \\
    20    & 16,781 &	2,479 &	2,500 & 16,914 & 5,817 & 950  \\
    22    & 21,064 &	2,768 &	2,500 & 43,517 & 8,976 & 1,124 \\
    24    & 53,108 &	39,808 &	2,500 & 53,518 & 24,885 & 1,520 \\
    26    & 69,995 &	43,514 &	2,500 & 70,005 & 23,064 & 1,578 \\
    28    & 90,080 &	68,198 &	2,500 & 90,080 & 26,807 & 1,640 \\
    30    & - & - & - & 74,998 & 23,374 & 1,471 \\
    32    & - & - & - & 74,023 & 27,337 & 1,524 \\
    34    & - & - & - & 82,445 & 32,506 & 1,622 \\
    36    & - & - & - & 91,031 & 44,523 & 1,698 \\
    \hline
    \shortstack{Geometric\\mean} & 9,995 & 1,845 & 1,350 & 18,138 & 4,043 & 649 \\
    \hline
    \end{tabular}
    \begin{tablenotes}\footnotesize
    \item[] \emph{Notes: A dash (``-'') is used to indicate when a solver did not find a feasible solution in the given limits.}
    \end{tablenotes}
    \end{threeparttable}
  \label{pxresults1}
\end{table*}

\begin{table*}[]
  \centering
  \caption{Computational results for $Q(x)/\Delta P(x)$ optimization--Part 2}
    \begin{threeparttable}
    \begin{tabular}{c|ccc|ccc}
    \hline
    \multirow{2}[0]{*}{\# of tubes} & \multicolumn{3}{c|}{glcSolve} & \multicolumn{3}{c}{MIDACO}  \\
          & $\frac{Q(x)}{\Delta P(x)}$ ($W/kPa$     & Time  & \shortstack{Function\\evaluations} & $\frac{Q(x)}{\Delta P(x)}$ ($W/kPa$     & Time  & \shortstack{Function\\evaluations} \\
    \hline
    4     & 413 & 4.40  & 5   & 413 & 1,323 & 2,500   \\
    6     & 280 & 35 & 37     & 280 & 512 & 2,500  \\
    8     & 1,443 & 387 & 345 & 1,417 & 300 & 2,500   \\
    10    & 8,894 & 960 & 531 & 8,905 & 721 & 2,500   \\
    12    & 8,216 & 1,553 & 654  & 2,944 & 196 & 2,500  \\
    14    & 26,219 & 3,186 & 775 & 26,219 & 400 & 2,500   \\
    16    & 24,393 & 10,939 & 1,299 & 24,295 & 1,023 & 2,500  \\
    18    & 30,830 & 8,234 & 1,104 & 30,767 & 1,391 & 2,500 \\
    20    & 16,914 & 6,759 & 963  &  -     & -     & -  \\
    22    & 43,517 & 10,581 & 1,134 &  -     & -     & -  \\
    24    & 53,646 & 27,617 & 1,551 & 35,930 & 5,641.30 & 2,500 \\
    26    & 70,005 & 26,683 & 1,577 &  -     & -     & - \\
    28    & 90,080 & 34,272 & 1,673 &  -     & -     & - \\
    30    & 75,109 & 37,244 & 1,590 &  -     & -     & - \\
    32    & 74,023 & 35,346 & 1,536 &  -     & -     & - \\
    34    & 82,445 & 77,147 & 1,641 &  -     & -     & - \\
    36    & 91,031 & 86,400 & 1,237 &  -     & -     & - \\
    \hline
    \shortstack{Geometric\\mean} & 18,141 & 4,813 & 643 & 5,248 & 768   & 2,500   \\
    \hline
    \end{tabular}
    \begin{tablenotes}\footnotesize
    \item[] \emph{Notes: A dash (``-'') is used to indicate when a solver did not find a feasible solution in the given limits.}
    \end{tablenotes}
    \end{threeparttable}
  \label{pxresults1}
\end{table*}

\begin{table*}[]
  \centering
  \caption{Computational results for $Q(x)/\Delta P(x)$ optimization--Part 3}
    \begin{threeparttable}
    \begin{tabular}{c|ccc}
    \hline
    \multirow{2}[0]{*}{\# of tubes} & \multicolumn{3}{c}{NOMAD} \\
          & $\frac{Q(x)}{\Delta P(x)}$ ($W/kPa$     & Time  & \shortstack{Function\\evaluations} \\
    \hline
    4     & 413	& 2 &	16  \\
    6     & 279 &	14 &	207 \\
    8     & 1,432 & 43 &	191 \\
    10    & 8,905 &	66 &	195 \\
    12    & 2,894 &	99 &	183 \\
    14    & 26,219	& 317 &	163 \\
    16    & 24,306 &	973 &	434 \\
    18    & 30,889 &	756 &	481 \\
    20    & -	& - & -\\
    22    & -	& - & -\\
    24    & 33,529 &	4,874 &	812 \\
    26    & 49,829	& 4,316	& 859 \\
    28    & 90,080	& 6,440	& 443 \\
    30    & 57,070	& 10,388	& 1,223 \\
    32    & 73,859	& 9,604	& 997 \\
    34    & -	& - & -\\
    36    & -	& - & -\\
    \hline
    \shortstack{Geometric\\mean} & 11,371 & 447 & 314 \\
    \hline
    \end{tabular}
    \begin{tablenotes}\footnotesize
    \item[] \emph{Notes: A dash (``-'') is used to indicate when a solver did not find a feasible solution in the given limits.}
    \end{tablenotes}
    \end{threeparttable}
  \label{pxresults2}
\end{table*}

Results for the optimization of the two objective functions showed that TOMLAB/glcDirect and TOMLAB/glcSolve can efficiently solve the proposed model and produce optimal or near-optimal solutions. Comparing those results with the complete enumeration results for the five heat exchangers with $4$ to~ $12$ tubes, TOMLAB/glcDirect and TOMLAB/glcSolve found:
\begin{itemize}
  \item For the optimization of heat capacity, four optimal solutions and a near-optimal solution that deviates from the optimal solution by only $0.05$\%
  \item For the optimization of the ratio of the heat capacity to the pressure difference across the heat exchanger, two optimal solutions and three near-optimal solutions that deviate from the optimal solution by an average of only $0.15$\%.
\end{itemize}
Hence, the use of constraint programming on the smaller heat exchangers verifies that the results generated by TOMLAB/glcDirect and TOMLAB/glcSolve are optimal or near-optimal.

\section{Conclusions}
\label{sec6}

Optimization of a heat exchanger design is a very important task since it can improve the performance of the designed heat exchanger. Most of the proposed methods aim to optimize the heat capacity by finding optimal values for structural parameters, such as tube thickness and fin spacing, and operating conditions, such as the refrigerant temperature and pressure. Another significant task when designing a highly efficient heat exchanger is to optimize the refrigerant circuitry. Design engineers currently choose the refrigerant circuitry according to their experience and heat exchanger simulations. However, there are many possible refrigerant circuitry candidates and thus, the design of an optimized refrigerant circuitry is difficult.

In this paper, we proposed a new formulation for the refrigerant circuitry design problem. We modeled this problem as a constrained binary optimization problem. We used CoilDesigner to simulate the performance of different refrigerant circuitry designs. CoilDesigner acts as a black-box since the exact relationship of the objective function with the decision variables is not explicit. DFO algorithms are suitable for solving this black-box model since they do not require explicit functional representations of the objective function and the constraints. We applied five DFO solvers on $17$ heat exchangers. Results showed that TOMLAB/glcDirect and TOMLAB/glcSolve can find optimal or near-optimal refrigerant circuitry designs on all instances.  We also used constraint programming methods to verify the results of the DFO methods for small heat exchangers.  The results show that the proposed method provides optimal refrigerant circuitries satisfying realistic manufacturing constraints.  The proposed heat exchanger circuitry optimization methods generate optimal or near-optimal circuit designs without requiring extensive domain knowledge.  As a result, the proposed approach can be readily applied to different types of heat exchangers.

Another contribution of the paper was the comparison between four mixed-integer constrained DFO solvers and one box-bounded DFO solver on industrially-relevant problems.  These solvers were applied to optimize heat exchanger circuitry using two different thermal efficiency criteria.  We found that TOMLAB/glcDirect and TOMLAB/glcSolve had the best performance.

In future work, we plan to consider other important performance metrics such as the shortest joint tubes and the production cost. In addition, future work should also optimize other parameters of the heat exchanger design, e.g., the tube thickness, the fin spacing, and the refrigerant temperature and pressure.

\Urlmuskip=0mu plus 1mu\relax

\end{singlespacing}

\end{document}